\documentclass[%
 reprint,
 showpacs,
 amsmath,amssymb,
 aps,
 prc,
]{revtex4-1}

\usepackage{graphicx}% Include figure files
\usepackage{dcolumn}% Align table columns on decimal point
\usepackage{bm}% bold math
\usepackage{lipsum}
\usepackage{csquotes}
\usepackage{textcomp}
\usepackage{nicefrac} 
\usepackage{color}

\newcommand{\ket}[1]{|#1\rangle}
\newcommand{\braket}[2]{\langle#1|#2\rangle}
\newcommand{\matrEL}[3]{\langle#1|#2|#3\rangle}

\newcommand{\leftsub}[2]{{\vphantom{#2}}_{#1}{#2}}

\begin{document}

\preprint{APS/123-QED}

\title{Deuteron-induced nucleon transfer reactions within  \\ an \textit{ab initio} framework: 
First application to p-shell nuclei}% Force line breaks with \\

\author{Francesco Raimondi}
\email[E-mail:~]{f.raimondi@surrey.ac.uk}
\altaffiliation[Present address:~]{Department of Physics, Faculty of Engineering and Physical Sciences, University of Surrey, Guildford GU2 7XH, United Kingdom.}
\affiliation{TRIUMF, 4004 Wesbrook Mall, Vancouver BC, V6T 2A3, Canada.}
\author{Guillaume Hupin}
\email[E-mail:~]{hupin@ipno.in2p3.fr}
\altaffiliation[Present address:~]{CEA, DAM, DIF, F-91297 Arpajon, France.}
\affiliation{Lawrence Livermore National Laboratory, P.O Box 808, L-414, Livermore, California 94551, USA}
\affiliation{Institut de Physique Nucl\'eaire, IN2P3-CNRS, Universit\'e Paris-Sud, F-91406 Orsay  Cedex, France.}
\author{Petr Navr\'atil}
\email[E-mail:~]{navratil@triumf.ca}
\affiliation{TRIUMF, 4004 Wesbrook Mall, Vancouver BC, V6T 2A3, Canada}
\author{Sofia Quaglioni}
\email[E-mail:~]{quaglioni1@llnl.gov}
\affiliation{Lawrence Livermore National Laboratory, P.O Box 808, L-414, Livermore, California 94551, USA.}

\date{\today}% It is always \today, today,

\begin{abstract}
\begin{description}
\item[Background] Low-energy transfer reactions in which a proton is stripped from a deuteron projectile and dropped into a target  play a crucial role in the formation of nuclei
in both primordial and stellar nucleosynthesis, as well as in the study of exotic nuclei using radioactive beam facilities and inverse kinematics. {\em Ab initio} approaches %, in which the reactant nuclei encompass all constituent nucleons as active degrees of freedom that interact \textit{via} nucleon-nucleon forces informed by quantum chromodynmaics, 
have been successfully applied to describe the $^3$H$(d,n)^4$He and $^3$He$(d,p)^4$He fusion processes. %deuteron collisions on 
%with $s$-shell targets.   %Experimentally such reactions are also widely used to study the single-particle nature of nuclear bound and resonant states, and have become particularly important in the investigation of exotic nuclei using new radioactive beam facilities and inverse kinematics. 
\item[Purpose]   An {\em ab initio} treatment of transfer reactions would also be desirable for heavier targets. %Deuteron collisions on $s$-shell targets have recently become accessible to {\em ab initio} approaches, leading to accurate calculations of structure and reaction observables  in which the reactant nuclei encompass all constituent nucleons as active degrees of freedom that interact \textit{via} nuclear forces grounded in the underlying theory of Quantum Chromodynamics.  
In this work, we extend the {\em ab initio} description of %the more difficult treatment of 
$(d,p)$ reactions to processes with light $p$-shell nuclei. As a first application, we study the elastic scattering of deuterium on $^7$Li and the ${}^{7}$Li($d$,$p$)${}^{8}$Li  transfer reaction based on a two-body Hamiltonian.
 
%investigate the reaction mechanism of the ${}^{7}$Li($d$,$p$)${}^{8}$Li transfer reaction, and describe the resonant part of the ${}^{9}$Be spectrum above the $d$+${}^{7}$Li threshold. %Calculate deuteron-nucleon transfer observables in p-shell nuclei within a framework in which the reactant nuclei encompass all constituent nucleons as active degrees of freedom, interacting \textit{via} Quantum Chromodynamics (QCD)-based microscopic nucleon-nucleon interactions. 
\item[Methods] %We seek the solutions of the nuclear many-body Sch\"odinger equation as expansions %of the composite nucleus 
%Starting from nucleon-nucleon interactions grounded in the underlying theory of quantum chromodynamics, 
We use the no-core shell model to compute the wave functions of the nuclei involved in the reaction, %as expansions in harmonic oscillator many-body states, according to the no-core shell model formalism. 
and describe the dynamics between targets and projectiles %nuclei is then described 
with the help of microscopic-cluster states in the spirit of the resonating group method. %We use the \lq calculable\rq  ~R-matrix method on Lagrange mesh to solve the ensuing one-dimensional integral-differential coupled channel equations, and compute the elements of the scattering matrix.  
\item[Results] The shape of the excitation functions for deuteron impinging on ${}^{7}$Li are qualitatively reproduced up to the deuteron breakup energy. The interplay between $d$-$^7$Li and $p$-$^8$Li particle-decay channels determines some features of the ${}^{9}$Be spectrum above the $d$+${}^{7}$Li threshold. %From an analysis of the resonant peaks in the %cross sections of ${}^{7}$Li($d$,$d$)${}^{7}$Li elastic and ${}^{7}$Li($d$,$p$)${}^{8}$Li  cross section, 
%transfer reactions, 
Our prediction for the parity of the 17.298 MeV resonance %at 0.60 MeV above the $d$+${}^{7}$Li threshold 
is at odds with the experimental assignment. %The dominance of the $p$-$^8$Li versus $d$-$^7$Li channel in this resonance could explain why   such $^9$Be state is hardly visible in the ${}^{7}$Li($d$,$d$)${}^{7}$Li  data.
%We study the reaction mechanism of the ${}^{7}$Li($d$,$p$)${}^{8}$Li transfer reaction and describe the resonant part of the ${}^{9}$Be spectrum above the $d$+${}^{7}$Li threshold. We discuss the spin-parity assignment of the
%resonant peaks in the cross sections of ${}^{7}$Li($d$,$d$)${}^{7}$Li elastic and ${}^{7}$Li($d$,$p$)${}^{8}$Li  transfer reactions.  
\item[Conclusions] Deuteron stripping reactions with $p$-shell targets can now be computed {\em ab initio}, but calculations are very demanding. A quantitative description of the ${}^{7}$Li($d$,$p$)${}^{8}$Li reaction %and of the resonant part of the ${}^{9}$Be spectrum above the $d$+${}^{7}$Li threshold 
will require further work to include the effect of three-nucleon forces and additional decay channels, and improve the convergence rate of our calculations.   %The shape of the excitation functions for deuteron impinging on ${}^{7}$Li are qualitatively reproduced up to the deuteron breakup energy. The parity of the calculated resonance at about 0.60 MeV above the $d$+${}^{7}$Li threshold is at odds with the experimental assignment.

\end{description}
\end{abstract}

\maketitle

%\tableofcontents

\section{\label{sec:level1}Introduction}
Since the introduction of the Born approximation by Stuart Thomas Butler in the Fifties~\cite{Butler1950}, theoretical studies of deuteron-induced one-nucleon stripping reactions have been advancing significantly %in the description of the reaction dynamics  
(see Ref.~\cite{Johnson2014} for a review
of recent developments and open problems on the topic). Such an effort has been motivated by the fact that transfer reactions
have become one of the prominent tool for nuclear structure investigations, in particular to extract spectroscopic information from nuclei. Still, the predictive capability of practical modern theories, %phenomenological models, 
relying on effective potentials and different approximations to treat the internal wave functions of the reactant nuclei and/or breakup of the deuteron, has been challenged by the 
advent of low-energy radioactive beams and the era of measurements of exotic phenomena related to astrophysical processes and nuclear structure away from the valley of stability~\cite{Jones2013}. 

The fact that the deuteron is a shallow bound state of a neutron and a proton plays a crucial role in the description of these transfer reactions. 
Three-body models, using as degrees of freedom a neutron, a proton (initially bound in the incident projectile) and a target nucleus, are well suited to account for the non-resonant continuum of the deuteron and its polarization effects below the breakup threshold. Examples include Faddeev-type calculations (e.g. Ref.~\cite{Deltuva2005}), adiabatic approaches~\cite{Johnson2014} and the Continuum Discretized Coupled Channel (CDCC) method (e.g. Ref.~\cite{Yahiro1986}).   However, questions remain on how to faithfully connect such models with the many-body problem which truly characterizes the reaction process.
%The non-resonant continuum of the deuteron and its polarization effects below the breakup threshold are taken into account exactly in Faddeev-type calculations (e.g. Ref.~\cite{Deltuva2005}), and alsoin three-body models~\cite{Johnson2014} and in the Continuum Discretized Coupled Channel (CDCC) method (e.g. Ref.~\cite{Yahiro1986}), based on different treatments of the proton-neutron part of the nuclear wave function. In three-body models this is implemented with a projected scattering wave functions which describes deuteron scattering and break-up exactly, whereas in the CDCC approach  \textit{via} a discretization of the corresponding continuous momentum variable describing their relative motion.
%When considering transfer reactions at low energy, as in the case of astrophysical processes taking place during the stellar burning, it is also crucial to take into account the short-range many-body correlations between the deuteron and the struck nucleus that lead to the formation of the compound system. This is due to the relatively small energy difference between the two thresholds, the one of the deuteron-target system and the one of the exit channel with the stripped nucleon. 
Indeed, to use the words of Butler~\cite{Butler1950}, in the low-energy regime typical of astrophysical processes %taking place during the stellar burning %The low-energy regime implies a lack of other significant open channels between these two thresholds, especially for light nuclei, meaning that the transfer process also is indeed probing probes the internal structure of the compound nucleus and its resonances. 
%Therefore,  to use the words of Butler in the pioneering work quoted above, this amounts to allowing our 
the theory has also to handle \enquote{The possibility that the whole deuteron may enter the nucleus}, giving rise to a complex interaction process with all the constituent nucleons in the target. More specifically this should be achieved by using realistic nuclear interactions and enforcing exactly the Pauli principle for a system of fermions. 

Corrections owing to %The importance of 
the full antisymmetrization of the nuclear wave function %for transfer reactions in the low-energy regime, 
have been explored in the context of three-body models%methods based on phenomenological potentials
~\cite{Johnson1982,Tostevin1987}, and were found to be important %. It is in fact 
for deuteron incident energies below the Coulomb and centrifugal barrier of the nucleus. There the adiabatic approximation of treating %where 
the proton-neutron distance 
%in the deuteron is treated 
as a \lq frozen\rq  \ parameter breaks down and %This means that 
the effects of the presence of the projectile in the nuclear interior are no longer negligible.

Another important many-body correction in a few-body description of the low-energy interactions of a projectile and a target are core excitations.  Excitations of the target nucleus
have been directly or indirectly accounted for in the CDCC~\cite{Summers2006} approach, adiabatic three-body models~\cite{Johnson2014_II} and, recently, in the distorted-wave Born approximation model~\cite{GomezRamos2015}. Important core excitations effects have also been found in Faddeev-type calculations for nuclear reactions~\cite{Deltuva2013}. Multiple core excitations are in particular needed
when deuteron stripping reactions populating resonance states of the final nucleus are considered, as it is the case for the 
CDCC extension to transfer reactions in the continuum~\cite{Mukhamedzhanov2011}.

In an \textit{ab initio} description, all of the above described aspects of the reaction mechanism should be addressed by considering all nucleons as active degrees of freedom that interact through all relevant [nucleon-nucleon (NN), three-nucleon (3N), \textit{etc}.] sectors of a realistic nuclear force, and by fully enforcing the Pauli principle. In this respect our method of choice is the no-core shell model (NCSM)~\cite{Navratil2009} combined with the resonating group method (RGM)~\cite{Wildermuth1977}. The NCSM/RGM~\cite{Quaglioni2009} approach relies on a projectile-target microscopic cluster ansatz for the $A$-nucleon wave function where the 
individual clusters, with ($A$-$a$) and $a$ nucleons ($a\leq A$), respectively, are eigenstates of their respective intrinsic Hamiltonians expanded in an harmonic oscillator (HO) basis~\cite{Barrett2013131} of the NCSM. For the dynamics among the nucleons, the NCSM/RGM employs realistic NN and 3N nuclear forces, that in last two decades have been connected 
to quantum chromo-dynamics (QCD) through chiral perturbation theory~\cite{Machleidt20111}. A natural extension of the NCSM/RGM formalism is to consider an enlarged model-space including NCSM eigenstates of the $A$-nucleons system, i.e. the
composite nucleus in the reaction. This extension, which we call no-core shell model with continuum (NCSMC)~\cite{Baroni2013a,Baroni2013b}, accelerates the convergence of the calculation by providing a more efficient description  of the short-range physics at the $A$-body level that is hard to capture within the cluster ansatz of the NCSM/RGM formalism. 

%Efforts are ongoing in order to include the three-body (3N) interaction in the description
%of nuclear reaction with light projectiles~\cite{Hupin2013,Hupin2015}.

%
\begin{figure}[t]
\centering
  \begin{tabular}{@{}cc@{}}
\includegraphics[width=.22\textwidth]{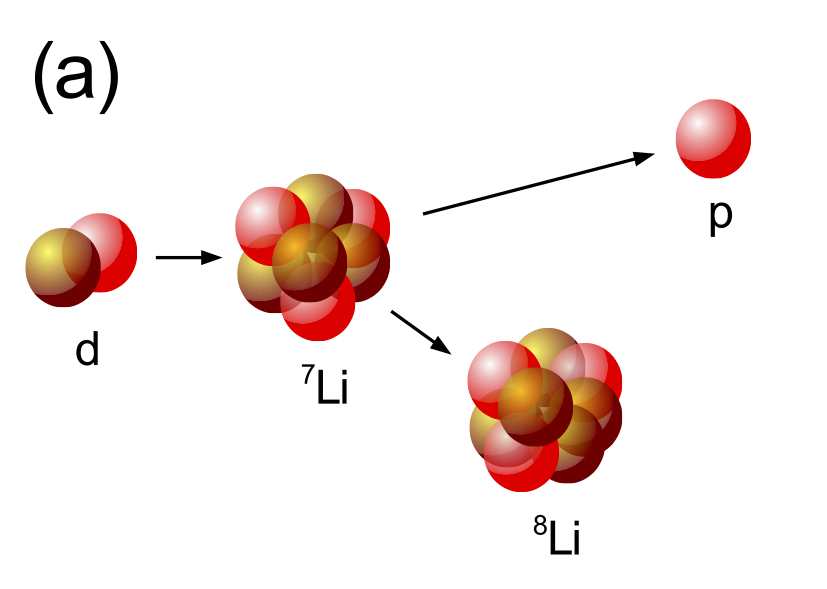}% Here is how to import EPS art
&
\includegraphics[width=.22\textwidth]{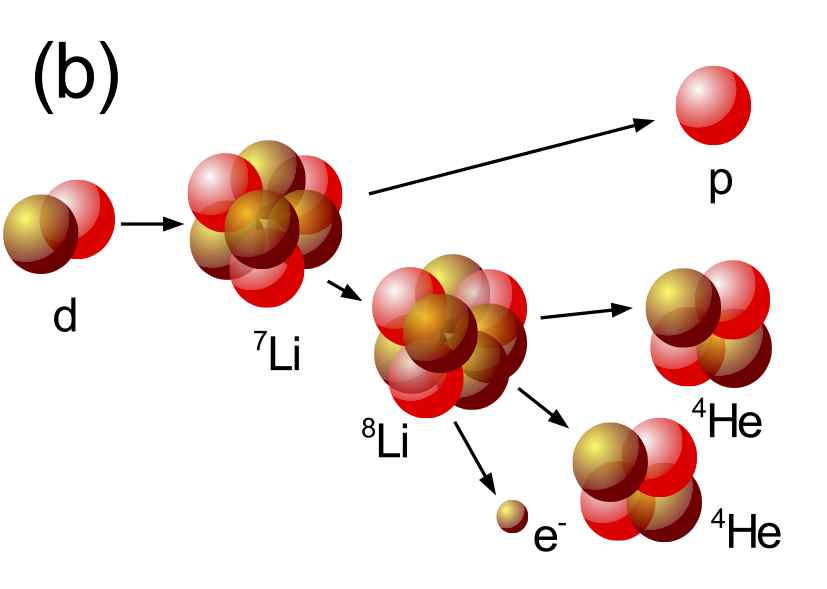}
\end{tabular}
\caption{\label{cartoons}${}^{7}$Li($d$,$p$)${}^{8}$Li reaction (a) as described in the present work, with the mass partitions in the entrance and exit channels modeled with cluster wave functions. The reaction as it is measured in experiment  (b), where the detection apparatus can be based on 1) the counting of protons, 2) the beta decay of $^8$Li or 3) the yield of the $^8$Be (not shown in the figure) delayed alphas following the $^8$Li beta decay.}
\end{figure}
The main purpose of the present paper is to extend  the NCSM/RGM description of $(d,N)$  transfer reactions introduced in Ref.~\cite{Navratil2012} to processes involving 
$p$-shell %($A$-2$>$4) 
targets. The much larger size of the model space compared to $s$-shell targets and the memory required for storing the Hamiltonian matrix elements had so far precluded such calculations.
These computational limitations %that had so-far precluded such calculations %introduced and first applied to the $^3$H$(d,n)^4$He and $^3$He$(d,p)^4$He fusion processes in Ref.~\cite{Navratil2012}, This is achieved 
are here overcome by generalizing to reactions with a deuteron projectile optimized algorithms already applied to the description of nucleon-nucleus scattering~\cite{Roth2014,Langhammer2015}. %that reduce the overhead of the calculation
%due to the huge size of the model space, and the memory required for storing the Hamiltonian matrix elements~\cite{Roth2014}.  The optimized algorithm has been already applied to the description of nucleon-nucleus scattering~\cite{Langhammer2015}, and we extend it here to reactions with a deuteron projectile.

As first interesting application, we compute the $^7$Li($d$,$p$)$^8$Li transfer reaction at energies below the deuteron
breakup threshold (see cartoons of Fig.~\ref{cartoons}). The inelastic and transfer scattering of deuteron on $^7$Li targets has been repeatedly measured in connection with the measurement of the radiative proton capture on $^7$Be %proton-capture reaction
~\cite{Parker1966,McClenahan1975,Elwyn1982,Filippone1982,Filippone1983,Strieder1996,Weissman1998678}. The main resonant peak at $\sim$0.60 MeV above the deuteron+$^7$Li threshold in the $^9$Be spectrum (corresponding to deuterons of $\sim$0.78 MeV kinetic energy), needs to be accurately measured in order to calibrate %the thickness of the $^7$Be target, i.e. 
the mean areal density of $^7$Be atoms in the targets used in the proton-capture measurement~\cite{Adelberger1998}. Moreover, the mechanism of destruction of $^7$Li through scattering of deuterons has been considered~\cite{Boyd1993} as a possible explanation for what is known as cosmic lithium depletion problem~\cite{Fields2011}, in particular in the context of non-standard (inhomogeneous) Big Bang Nucleosynthesis models~\cite{Malaney1993}. 

%In the present work we focus on the NCSM/RGM formalism for the description of $(d,N)$ reactions, which is currently available only for $s$-shell targets %($A$-2$\leq$4) 
%due to computational limitations~\cite{Navratil2012}. 
We also present an initial set of NCSMC results for the ${}^{7}$Li($d$,$d$)${}^{7}$Li elastic reaction, where ${}^{9}$Be eigenstates are included in the model-space. The treatment of deuteron stripping reactions within the NCSMC framework is beyond the scope of the present work and will be presented elsewhere.

The content of the paper is as follows. In Section~\ref{sec:formalism} we revisit the main features of the NCSM/RGM basis, and describe the computation of the NCSM/RGM Hamiltonian kernels for deuteron transfer processes by means of an optimized algorithm. The more general formalism of the NCSM/RGM and NCSMC approaches is presented in Appendix~\ref{App_gen}, while some useful algebraic expressions for the Hamiltonian kernels in the case of deuteron-induced reactions are 
collected in Appendix~\ref{App_2}.
We show in Section~\ref{sec:Results} the calculated eigenphase shifts, $p-^8$Li and $d-^7$Li elastic phase shifts, and cross sections for the elastic ${}^{7}$Li($d$,$d$)${}^{7}$Li and transfer ${}^{7}$Li($d$,$p$)${}^{8}$Li processes. We then discuss some features of the  $^9$Be
energy spectrum above the $d$+$^7$Li threshold. Finally, conclusions are drawn in Section~\ref{sec:conclusion}.

\section{\label{sec:formalism}Formalism}
 
The \textit{ab initio} NCSM/RGM formalism was introduced in Ref.~\cite{Quaglioni2009} for the description of nucleon-nucleus collisions. The formalism was later extended in order to address deuteron impinging on a target~\cite{Navratil2011} and $(d,N)$ fusion -- or transfer -- reactions with an $s$-shell target~\cite{Quaglioni2012,Navratil2012}. The latter reaction mechanism is characterized by different mass partitions in the entrance and exit channels.  

The microscopic $A$-body wave function is cast in the form of a partial wave decomposition on spin channels constructed by applying appropriate antisymmetrization operators to product states of the internal wave functions of the clusters,
\begin{eqnarray}
|\Psi^{J^\pi T}\rangle &=&  \sum_{\nu} \int dr \,r^2\frac{g^{J^\pi T}_{\nu}(r)}{r}\,\hat{\mathcal A}_{\nu}\,|\Phi^{J^\pi T}_{\nu r}\rangle\ 
\label{eq:trial}.
\end{eqnarray}
The unknown relative-motion amplitudes, denoted by  $g^{J^\pi T}_{\nu}(r)$, depend on the auxiliary variable $r$, and $J$, $\pi$, and $T$ are the partial-wave angular momentum, parity and isospin.   The index $\nu=\{A_t  \alpha_t I_t^{\pi_t}T_t; A_p \alpha_p I_p^{\pi_p}T_p; s\ell\}$ [with $A_t=A-a$, $A_p=a$, and $I_{p(t)}$, ${\pi_{p(t)}}$, $T_{p(t)}$, $\alpha_{p(t)}$, $s$, and $\ell$ denoting, respectively, the projectile (target) angular momentum, parity, isospin, and energy quantum numbers, the channel spin and the relative orbital angular momentum] runs over the set of all the possible channels included in the calculation. For the ${}^{7}$Li($d$,$p$)${}^{8}$Li transfer reaction, these include both the $d$-$^7$Li entrance and the $p$-$^8$Li exit channels. %In a binary-cluster representation of the wave function this is $\nu=\{A_t  \alpha_t I_t^{\pi_t}T_t; A_p \alpha_p I_p^{\pi_p}T_p; s\ell\}$, with $A_t=A-a$, and $A_p=a$, and $I_{p(t)}$, ${\pi_{p(t)}}$, $T_{p(t)}$, $\alpha_{p(t)}$, $s$, and $\ell$ denoting, respectively, the projectile (target) angular momentum, parity, isospin, and energy quantum numbers, the channel spin and the relative orbital angular momentum.
The auxiliary variable $r$ in Eq.~(\ref{eq:trial}) is introduced using a Dirac delta $\delta(r{-}r_{A{-}a,a})$ in order to remove the dependence on  the inter-cluster relative coordinate $\vec r_{A-a,a}=r_{A-a,a}\hat r_{A-a,a}$ from the relative-motion amplitudes between the colliding clusters. This formal step allows the antisymmetrization operator $\hat{\mathcal A}_{\nu}$ to act only on the channel states %product state of the clusters. For the case of the binary channel we have,
 \begin{eqnarray}
|\Phi^{J^\pi T}_{\nu r}\rangle &=& \Big [ \big ( \left|A_t\, \alpha_t I_t^{\,\pi_t} T_t\right\rangle \left |A_p\,\alpha_p I_p^{\,\pi_p} T_p\right\rangle\big ) ^{(s T)}\nonumber\\
&&\times\,Y_{\ell}\left(\hat r_{A-a,a}\right)\Big ]^{(J^\pi T)}\,\frac{\delta(r-r_{A-a,a})}{rr_{A-a,a}},
\label{basis}
\end{eqnarray}
where $\left|A_{t(p)}\, \alpha_{t(p)} I_{t(p)}^{\,\pi_{t(p)}} T_{t(p)}\right\rangle$ are translational-invariant eigenstates of the target (projectile) %the states containing $A{-}a$ and $a$ nucleons ($a{<}A$) describe, respectively, a target $t$ and a projectile $p$ in a state of relative motion.
obtained within the NCSM %The internal degrees of freedom of the target and projectile are obtained 
\textit{via} a variational calculation in a $N_{\rm max}$-restricted HO many-body space with frequency $\hbar\Omega$~\cite{Navratil2009}. 

The binary channel states of Eq.~(\ref{basis}) are employed to compute the matrix elements of the relative kinetic energy operator $T_{\rm rel}(r)$ and all other non-localized operators entering the expressions of the NCSM/RGM kernels of Eqs.\eqref{N-kernel} and \eqref{H-kernel}, including the Coulomb interaction $\bar{V}_{\rm C}(r)$ [see Eq.~(\ref{Hamiltonian}) for the expression of the internal $A$-nucleon microscopic Hamiltonian].  On the other hand, for the localized terms arising from the non-identical permutations of nucleons pertaining to different clusters, it is convenient to expand the Dirac delta of Eq.~\eqref{basis} in a basis of HO radial functions with the same frequency $\hbar \Omega$ as the one describing the internal motion of the clusters,
\begin{eqnarray}
|\Phi^{J^\pi T}_{\nu r}\rangle & = & \sum_{n} R_{n\ell}(r, b)\, |\Phi^{J^\pi T}_{\nu n,b}\rangle\  \nonumber\\
&=& \sum_{n} R_{n\ell}(r, b) \Big [ \big ( \left|A_t\, \alpha_t I_t^{\,\pi_t} T_t\right\rangle \left |A_p\,\alpha_p I_p^{\,\pi_p} T_p\right\rangle\big ) ^{(s T)}\,\nonumber\\
&&\times Y_{\ell}\left(\hat \eta_{A-a}\right)\Big ]^{(J^\pi T)}\, R_{n\ell}(r_{A-a,a},b)\,\label{ho-basis-n}.
\end{eqnarray}
Here the Jacobi coordinate $\vec \eta_{A-a}$ proportional to the relative position between the centers-of-mass (c.m.) of the two clusters is defined as,
\begin{equation}
\vec \eta_{A-a} = \sqrt{\frac{(A - a) a}{A}} \left[ \frac{1}{A - a}\sum_{i = 1}^{A - a} \vec r_i - \frac{1}{a} \sum_{j =  A - a + 1}^A \vec r_j\right]\,,\label{jacobi2}
\end{equation}
and a dependence on the oscillator-length parameter  $b = \sqrt{A / [(A{-}a) a]} \sqrt{ \hbar / m \Omega}$ has been introduced.

Due to the increasing complexity in the antisymmetrization of translationally invariant wave functions for increasing number of particles, it is convenient to use a single-particle Slater determinant (SD) representation for the target states. In the NCSM, such SD eigenstates 
% The spurious c.m. motion exactly factorizes in a state of zero angular momenta and node. Accordingly, the frequency of the HO basis is $\hbar \Omega$. 
are given by the product of %related to 
the translationally invariant ones %through 
with the $0\hbar\Omega$ HO wave function of the target c.m. %\ coordinate of the cluster.
In the case of %For the target states in 
the ${}^{7}$Li($d$,$p$)${}^{8}$Li transfer reaction %with the total wave function of Eq.~(\ref{eq:schematic}), 
we have
\begin{eqnarray}
 | ^7\textrm{Li} \rangle_{\rm SD} \equiv | 7\, \alpha_{7} I_7^{\,\pi_7} T_7 \rangle\varphi_{00} (R_{\rm c.m.}^{(^7\textrm{Li})})\,, %&=&\left|^7\textrm{Li}, \, \alpha_{7} I_7^{\,\pi_7} T_7\right\rangle_{\rm SD}\nonumber \\ 
%& \equiv & | ^7\textrm{Li} \rangle_{\rm SD},
\label{target_1}
\end{eqnarray}
%\begin{eqnarray}
%   \left |A{=}2\,\alpha_2 I_2^{\,\pi_2} T_2\right\rangle \equiv  | d \rangle  ,
%\label{proj_1}
%\end{eqnarray}
belonging to the entrance channel together with the eigenstates of the deuterium $| d \rangle\equiv\left |A_p{=}2\,\alpha_2 I_2^{\,\pi_2} T_2\right\rangle$, and 
\begin{eqnarray}
| ^8\textrm{Li} \rangle_{\rm SD}\equiv| 8  \, \alpha_{8} I_8^{\,\pi_8} T_8\rangle \varphi_{00} (R_{\rm c.m.}^{(^8\textrm{Li})}) \,,% &=& \left| ^8\textrm{Li}, \, \alpha_{8} I_8^{\,\pi_8} T_8\right\rangle_{\rm SD}\nonumber \\ 
%&\equiv &| ^8\textrm{Li} \rangle_{\rm SD},
\label{target_2}
\end{eqnarray}
which is the remnant nucleus in the exit channel along with the scattered proton $| p \rangle\equiv\left |1\,\textrm{\textonehalf}\, \textrm{\textonehalf}\right\rangle$. 
%In Eqs.~(\ref{target_1},\ref{proj_1},\ref{target_2}) $ I_i^{\,\pi_i} T_i$ for $i$=2,7 and 8 denote intrinsic spin, parity and isospin of the states, that are ordered in energy according to the index $\alpha_i$. 
%In Eq.~(\ref{eq:schematic}) the coupling in square brackets implies the sum of the spins and isospins of $^7$Li($^8$Li) and $d$($p$) to total channel spin $s$($s'$) and isospin $T$($T'$), and the coupling of $s$($s'$) to the relative-motion orbital angular momentum $l$($l'$). 
Correspondingly, it is convenient to introduce  %The expressions corresponding to $|\Phi^{J^\pi T}_{\nu n,b}\rangle$ of Eq.~(\ref{ho-basis-n}) for the 
SD channel states according to (omitting the explicit reference to the HO length parameter),
%\begin{eqnarray}
%|\Phi^{J^\pi T}_{\nu r}\rangle &=& \Big [ \big ( \left|A{-}a\, \alpha_1 I_1^{\,\pi_1} T_1\right\rangle \left |a\,\alpha_2 I_2^{\,\pi_2} T_2\right\rangle\big ) ^{(s T)}\nonumber\\
%&&\times\,Y_{\ell}\left(\hat r_{A-a,a}\right)\Big ]^{(J^\pi T)}\,\frac{\delta(r-r_{A-a,a})}{rr_{A-a,a}}\,,\label{basis}
%\end{eqnarray}
\begin{eqnarray}
|\Phi^{J^\pi T}_{\nu n}\rangle_{\rm SD}   &=&    \Big [\big (   | ^7\textrm{Li} \rangle_{\rm SD}  | d \rangle  \big )^{(s T)}  Y_{\ell}(\hat R^{(d)}_{\rm c.m.})\Big ]^{(J^\pi T)}\nonumber \\ 
&& \times 
R_{n\ell}(R^{(d)}_{\rm c.m.})\,,
\label{SD-basis_1}
%\end{eqnarray}
%\begin{eqnarray}
\\
|\Phi^{J^\pi T}_{\nu' n'}\rangle_{\rm SD}   &=&    \Big [\big (   | ^8\textrm{Li} \rangle_{\rm SD}  | p \rangle  \big )^{(s' T)}  Y_{\ell'}(\hat r_A)\Big ]^{(J^\pi T)}\nonumber \\ 
&& \times R_{n'\ell'}(r_A)\,,
\label{SD-basis_2}
\end{eqnarray}
where  $\vec R^{(d)}_{\rm c.m.}$($\vec r_A$) is the coordinate of the deuterium (proton) projectile, and we now explicitly separate the two channels with different mass partition, i.e., $\nu=\{7 \, \alpha_7 I_7^{\pi_7}T_7; 2 \alpha_2 I_2^{\pi_2}T_2; s\ell\}$, and $\nu' = \{ 8 \, \alpha_8 I_8^{\pi_8} T_8; \, 1\,  \textrm{\textonehalf} \,\textrm{\textonehalf};\, s'\ell'\}$. 

The pairs of coordinates $\{\vec R^{(d)}_{\rm c.m.},\vec{R}_{\rm c.m.}^{(^7{\rm Li})}\}$ and $\{\vec r_A,\vec{R}_{\rm c.m.}^{(^8{\rm Li})}\}$ are orthogonal transformations of  the c.m.\ coordinate and relative coordinate $\vec \eta_{A-a}$ of the $A$-nucleon system. As a consequence, the SD channel states of Eqs.~\eqref{SD-basis_1} and \eqref{SD-basis_2} can be transformed into expansions on HO wave functions depending on these latter coordinates, with coefficients given by generalized HO brackets for two particles with mass ratio $\frac{a}{A-a}$. 
The spurious motion of the $A$-nucleon c.m.\ coordinate can then be exactly removed at the level of matrix elements of translationally invariant operators, such as the microscopic Hamiltonian. Such a procedure is described in detail in Section IIC of Ref.~\cite{Quaglioni2009}. Therefore, in the case of an $N_{\rm max}$ scheme HO basis, this simple transformation mixing the spurious c.m. and the relative motion of the colliding nuclei allows us to recover the fully translationally invariant NCSM/RGM kernels and to take advantage of the computationally efficient SD formulation of the target states.

%implies in fact the dependence on the relative coordinate between the clusters and the spurious c.m. coordinate of the target, and consequently of the entire $A$-nucleons system (see Sec. IIC in Ref.~\cite{Quaglioni2009} for the description of the procedure of removal of c.m. spuriousity at the level of matrix elements of translationally invariant operators).

The expressions in Eqs.~(\ref{SD-basis_1}) and \eqref{SD-basis_2} can be further worked out  to recast the projectile wave function too
as product of single-particle functions. For the $p$-$^8$Li channel this manipulation  reads~\cite{Quaglioni2009,Hupin2013},
%\begin{widetext}
\begin{align}
|\Phi^{J^\pi T}_{\nu' n'}\rangle_{\rm SD}   &%&=&    \Big [\big(  | ^8\textrm{Li} \rangle_{\rm SD} 
%\left |1\,\textrm{\textonehalf} \textrm{\textonehalf}\right\rangle\big )^{(s' T)}Y_{\ell'}(\hat{r}_{A})\Big ]^{(J^\pi T)}%\nonumber \\
%&& 
%R_{n'\ell'}(r_{A})
%\nonumber\\
%&&
=  \sum_j (-1)^{I_8+J+j}\left\{ \begin{array}{@{\!~}c@{\!~}c@{\!~}c@{\!~}} 
I_8 & \frac{1}{2} & s' \\[2mm] 
\ell' & J & j 
\end{array}\right\}  \hat{s'}\hat{j} \nonumber \\
%&&
%& \times \Big [ | ^8\textrm{Li} \rangle_{\rm SD}  
%\varphi_{n' \ell' j \frac12} (\vec{r}_A \sigma_A \tau_A)\Big]^{(J^\pi T)}
%\,,
&\times\sum_{M_8 m_j} \sum_{M_{T_8} m_t} 
 \left( \begin{array}{c c|c}
  I_8 & j & J \\
  M_8 & m_j & M_J
 \end{array} \right) \nonumber\\
&\times 
 \left( \begin{array}{c c|c}
  T_8 & \frac{1}{2} &  T \\
  M_{T_8} & m_t & M_{T}
 \end{array} \right)
  \left|^8\textrm{Li}, \, \alpha_8 I_8  M_8 T_8  M_{T_8}\right\rangle_{\rm SD} \nonumber\\[2mm]
 &\times \left|n \ell j m_{j} \textrm{\textonehalf}\, m_{t}\right\rangle
\label{SD-basis-SNP_1}
\end{align}
where %$\nu' = \{ 8 \, \alpha_8 I_8^{\pi_8} T_8; \, 1\,  \textrm{\textonehalf} \textrm{\textonehalf};\, s'\ell'\}$ and 
%$\varphi_{n' \ell' j m \frac12 m_t} (\vec{r}_A \sigma_A \tau_A)=$
%$\langle \vec{r}_A \sigma_A \tau_A | n \ell j m_{j} \frac12 m_{t} \rangle$ 
$ | n \ell j m_{j} \frac12 m_{t} \rangle$ is the HO single-particle wave function of the proton projectile and we used the notation $ \left( \begin{array}{c c|c}
  J_1 & J_2 & J \\
  M_1 & M_2 & M_J
 \end{array} \right) $ for Clebsch-Gordan coefficients. %, given by $R_{n'\ell'}(r_{A})$ $\big (Y_{\ell'}(\hat{r}_{A}) \chi_{\frac12}(\sigma_A)\big )^{(j)}_m \chi_{\frac12 m_t}(\tau_A)$.  
For the $d$-$^7$Li channel the manipulation is somewhat more involved but straightforward,  requiring angular
 momentum recoupling coefficients and the use of HO brackets~\cite{Navratil2011}, %In the following we specialize to the mass partition ($^7$Li,$d$) the general result of the recoupling of the many-body wave function introduced in Ref.
\begin{eqnarray}
|\Phi^{J^\pi T}_{\nu n}\rangle_{\rm SD}   &=&  \sum 
\left\{ \begin{array}{@{\!~}c@{\!~}c@{\!~}c@{\!~}} 
I_7 & I_2 & s \\[2mm] 
\ell & J & j 
\end{array}\right\}   
\left\{ \begin{array}{@{\!~}c@{\!~}c@{\!~}c@{\!~}} 
\ell & L_{ab} & {\ell}_2 \\[2mm] 
 s_2 & I_2 & I 
\end{array}\right\}
\left\{ \begin{array}{@{\!~}c@{\!~}c@{\!~}c@{\!~}} 
{\ell}_a & {\ell}_b & L_{ab} \\[2mm] 
 \frac{1}{2} & \frac{1}{2} & s_2 \\[2mm] 
 j_a & j_b & I 
\end{array}\right\}  
\nonumber\\[2mm] 
&&\times 
(-1)^{I_7+J+\ell+\ell_2+T_2} \, \hat{s}\, \hat{I}\, \hat{I}_2\, \hat{s}_2 \, \hat{j}_a\, \hat{j}_b\, \hat{L}_{ab}^2
\nonumber\\[2mm]
&&\times
 \left\langle n_a {\ell}_a n_b {\ell}_b L_{ab} \left. \right| n \ell n_2 {\ell}_2 L_{ab}\right\rangle_{d=1}
\nonumber \\[2mm] 
&& \times
\left \langle n_2 {\ell}_2 s_2 I_2 T_2 \left. \right| 2\,\alpha_2 I_2 T_2\right\rangle \; |\Phi^{J^\pi T}_{\kappa_{ab}}\rangle_{\rm SD}\,.
\label{SD-basis-SNP}
\end{eqnarray}
Here the sum runs over the quantum numbers $n_2,\ell_2, s_2$, $n_a, \ell_a, j_a$, $n_b, \ell_b, j_b$, $L_{ab}$, $j$, and $I$,
$\langle n_2 \ell_2 s_2 I_2 T_2 | 2\,\alpha_2 I_2 T_2\rangle $ are the coefficients of the projectile wave function expanded in the relative-coordinate HO basis, $\hat{s}=\sqrt{2s{+}1}$ etc., and $\langle n_a {\ell}_a n_b {\ell}_b L_{ab} | n \ell n_2 {\ell}_2 L_{ab}\rangle_{d=1}$ indicates an HO bracket for two particles with identical masses. In addition, we introduced  the cumulative quantum number $\kappa_{ab} \equiv \{7 \, \alpha_7 I_7 T_7$; $n_a \ell_a j_a \tfrac12; n_b \ell_b j_b \tfrac12; I T_2\}$ and the new SD channel states
%\begin{align}
%|\Phi^{J^\pi T}_{\kappa_{ab}}\rangle_{\rm SD} &= \Big [ | ^7\textrm{Li} \rangle_{\rm SD} 
%\left(\varphi_{n_a \ell_a j_a \frac12} (\vec{r}_A \sigma_A \tau_A) \right.
%\nonumber \\
%&\quad\phantom{=}\times \left. \varphi_{n_b \ell_b j_b \frac12} (\vec{r}_{A-1} \sigma_{A-1} \tau_{A-1})\right)^{(I T_2)}\Big ]^{(J^\pi T)}.
%\label{SD-basis-ab}
%\end{align}
%
%The basis states defined by the right-hand side of Eqs.~(\ref{SD-basis-SNP_1}) and (\ref{SD-basis-ab}) are now decoupled and expressed as a sum of components of the spins and isospins of the two clusters. In this basis we can take advantage of the second quantization formalism, which permits us to efficiently compute matrix elements of operators. To show this, we will explicitly derive the expression of the matrix elements of the intercluster interaction such that the essential quantity -a product of annihilation and creation operators averaged on the corresponding vacuum- is spelled out.
%
%According
%to the definition $\varphi_{n_a \ell_a j_a \frac12} (\vec{r}_A \sigma_A \tau_A) \equiv \langle \vec{r}_A \sigma_A \tau_A | n_a \ell_a j_a m_{j_a} \frac12 m_{t_a} \rangle$, the decoupling of the binary-cluster channel states in both spin and isospin %spaces reads [we use the notation $ \left( \begin{array}{c c|c}
%  J_1 & J_2 & J \\
%  M_1 & M_2 & M_J
% \end{array} \right) $ for Clebsch-Gordan coefficients], 
\begin{align}
	|\Phi^{J^\pi T}_{\kappa_{ab}}\rangle_{\rm SD} &= \sum_{M_7 M_I} \sum_{M_{T_7} M_{T_2}} \sum_{m_{j_a} m_{j_b}} \sum_{m_{t_a} m_{t_b}} 
 \left( \begin{array}{c c|c}
  I_7 & I & J \\
  M_7 & M_I & M_J
 \end{array} \right)  \nonumber\\
 &\times
 \left( \begin{array}{c c|c}
  T_7 & T_2 &  T \\
  M_{T_7} & M_{T_2}  & M_{T}
 \end{array} \right)   \left( \begin{array}{cc|c}
  j_a & j_b &  I \\
  m_{j_a}  & m_{j_b}  & M_{I}
 \end{array} \right) \nonumber \\
 & \times   \left( \begin{array}{c c|c}
 \frac{1}{2}& \frac{1}{2} &  T_2 \\
  m_{t_a}  & m_{t_b}  & M_{T_2}
 \end{array} \right) 
  \left|^7\textrm{Li}, \, \alpha_7 I_7  M_7 T_7  M_{T_7}\right\rangle_{\rm SD} 
  \nonumber\\[2mm]
 &\times
  \left|n_a \ell_a j_a m_{j_a} \textrm{\textonehalf}\, m_{t_a}\right\rangle  \left|n_b \ell_b j_b m_{j_b} \textrm{\textonehalf}\, m_{t_b}\right\rangle 
\label{SD-basis-ab}.
\end{align}

The basis states %defined by the right-hand side 
of Eqs.~(\ref{SD-basis-SNP_1}) and (\ref{SD-basis-ab}) are now expressed in terms of uncoupled products of single-particle states. %a sum of components of the spins and isospins of the two clusters. In this basis we can 
This allows us to take advantage of the second quantization formalism %, which permits us to 
and efficiently compute matrix elements of operators. %To show this, here we derive the expression of the matrix elements of the intercluster interaction such that the essential quantity -a product of annihilation and creation operators averaged on the corresponding vacuum- is spelled out.
%
%The goal is now to compute the matrix elements of the nuclear interaction with respect to the wave functions in the decoupled form, Eqs.~(\ref{SD-basis-decoupled-1}) and (\ref{SD-basis-decoupled-1onenucleon}). In particular the matrix elements of the interaction between nucleons belonging to different clusters, referred also as coupling kernels, play a crucial role in describing the transfer reaction process. The form of the coupled-channel many-body Schr\"odinger equation in the NCSM/RGM framework is revisited in Appendix~\ref{App_1}, whereas the explicit form of the coupling kernels required for the correct description of the deuteron-induced transfer reactions are shown in Appendix~\ref{App_2}.
Among the components of the Hamiltonian kernel of Eqs.~\eqref{V-kernel} and \eqref{pot-coupled}, three are especially demanding in terms of the required computational resources because they involve operations on more than two nucleons of the target. The first one, appearing in %the last term of the potential kernel in 
Eq.(\ref{antisym_HAm2}), is a term diagonal in the ($^7$Li,$d$) mass partition and depends on a three-body density matrix of the target nucleus. 
Adopting  the notation $\langle   a b |   V |  c d  \rangle$ for the antisymmetrized two-nucleon potential matrix elements, its explicit expression is 
\begin{widetext}
 \begin{eqnarray}  
\label{Summary_II}
&& \left<\Phi_{k'_{a b}}^{J^\pi T}\right| \left( V_{A,A-4}  \hat P_{A-2,A-1} \hat P_{A-3,A} \right)\left|\Phi_{k_{a b}}^{J^\pi T}\right> = \nonumber \\
&&\sum_{M'_7 M'_{I'}} \sum_{M'_{T'_7} M'_{T'_2}} \sum_{m'_{j'_a} m'_{j'_b}} \sum_{m'_{t'_a} m'_{t'_b}} 
 \left( \begin{array}{c c|c}
  I'_7 & I' & J \\
  M'_7 & M'_{I'} & M_J
 \end{array} \right)  
 \left( \begin{array}{c c|c}
  T'_7 & T'_2 &  T \\
  M'_{T'_7} & M'_{T'_2}  & M_{T}
 \end{array} \right)
  \left( \begin{array}{c c|c}
  j'_a & j'_b &  I' \\
  m'_{j'_a}  & m'_{j'_b}  & M'_{I'}
 \end{array} \right) \left( \begin{array}{c c|c}
 \frac{1}{2}& \frac{1}{2} &  T'_2 \\
  m'_{t'_a}  & m'_{t'_b}  & M'_{T'_2}
 \end{array} \right) \nonumber \\
&& \sum_{M_7 M_I} \sum_{M_{T_7} M_{T_2}} \sum_{m_{j_a} m_{j_b}} \sum_{m_{t_a} m_{t_b}} 
 \left( \begin{array}{c c|c}
  I_7 & I & J \\
  M_7 & M_I & M_J
 \end{array} \right)  
 \left( \begin{array}{c c|c}
  T_7 & T_2 &  T \\
  M_{T_7} & M_{T_2}  & M_{T}
 \end{array} \right)
  \left( \begin{array}{c c|c}
  j_a & j_b &  I \\
  m_{j_a}  & m_{j_b}  & M_{I}
 \end{array} \right)  \left( \begin{array}{c c|c}
 \frac{1}{2}& \frac{1}{2} &  T_2 \\
  m_{t_a}  & m_{t_b}  & M_{T_2}
 \end{array} \right) \nonumber \\
 &&  \sum_{\beta \gamma \delta} \frac{1}{2(A-4)(A-3)(A-2)}\   \leftsub{\rm SD}{\left\langle ^7\textrm{Li}  , \Omega'_7 \left| \hat{a}^{\dagger}_{\beta}   \hat{a}^{\dagger}_{a}   \hat{a}^{\dagger}_{b} \hat{a}_{b'} \hat{a}_{\delta} \hat{a}_{\gamma}  \right|{}^7\textrm{Li} , \Omega_7 \right\rangle}_{\rm SD}\  \left\langle   \beta, a' \left|   V \right|  \gamma\delta  \right\rangle.
 \end{eqnarray}
 \end{widetext}
For sake of generality we %kept in the expression 
use $A$ to indicate the total number of particle in the system, which in our case is 
 $A=9$. In addition, here and in the following equation %the expression of Eq.~(\ref{Summary_II}) and in the following Hamiltonian kernels 
 we label the single-particle states of nucleons
which appear in the wave functions of the projectile with Latin letters as before, whereas we use Greek letters for those appearing in the expansion of the target nucleus wave function. % are labeled with Greek letters. 
The capital $\Omega_i$ ($i$=7,8) is instead reserved for the quantum numbers of the target states ($\Omega_i \equiv I_i  M_i T_i  M_{T_i}  $). By introducing coupled densities and performing further algebraic manipulations, Eq.~(\ref{Summary_II}) can be cast into a coupled form and one recovers Eq.~(24) of Ref.~\cite{Navratil2011} and Eq.~(\ref{V-kernel_5-compl}) in Appendix~\ref{App_2}.
 
The other two components (see last two Hamiltonian coupling kernels in Eq. (19) of Ref.~\cite{Quaglioni2012} and Eq.~(\ref{pot-coupled-last}) in Appendix~\ref{App_2}) appear in the coupling kernels between the ($^7$Li,$d$) and ($^8$Li,$p$) mass partitions and depend on a density matrix which contains 
two creation and three annihilation operators. Hamiltonian kernels which have one unpaired creation or annihilation operator correspond to the one-nucleon transfer 
part of the scattering process, where the final nucleus contains the stripped nucleon from the projectile. %On the other hand, since we aim at describing the dynamics of the scattering process by including all the relevant
%channels, we include in our calculations of the transfer reaction also the Hamiltonian kernels of the \lq elastic\rq \ deuteron-nucleus scattering, such as the one depending
%of the three-body density written above.
 For reasons of computational efficiency (it is easier to produce the list of all possible triplets of annihilation operators acting on a given many-body state, than produce the list of creation operators that must be compatible with both initial and final states), we cast these kernels in such a way that three annihilation operators and two creation ones are displayed in the density matrices, yielding %These kernels are given by a sum of two contributions (see last two Hamiltonian coupling kernels kernels in Eq. (19) of Ref.~\cite{Quaglioni2012} and in Eq.~(\ref{pot-coupled}) in Appendix~\ref{App_2}),
\begin{widetext}
   \begin{eqnarray}  
\label{Summary_I_bis_total}
&&    \left<\Phi_{k'_{a}}^{J^\pi T}\right|  \frac{1}{2}\hat P_{A-2,A} V_{A-3,A-2} + V_{A-3,A-2}   \hat P_{A-2,A}   \left|\Phi_{k_{a b}}^{J^\pi T}\right> =  \nonumber \\
&& \frac{(-1)^{A}}{2(A-3)(A-2)\sqrt{A-1}} \sum_{M'_8 m'_{j'_a}} \sum_{M'_{T_8} m'_{t'_a}} 
 \left( \begin{array}{c c|c}
  I'_8 & j'_a & J \\
  M'_8 & m'_{j'_a} &  M_J
 \end{array} \right)  
 \left( \begin{array}{c c|c}
  T'_8 & \frac{1}{2} &  T \\
  M'_{T_8} & m'_{t'_a} & M_{T}
 \end{array} \right) \nonumber \\
 && \sum_{M_7 M_I} \sum_{M_{T_7} M_{T_2}} \sum_{m_{j_a} m_{j_b}} \sum_{m_{t_a} m_{t_b}} 
 \left( \begin{array}{c c|c}
  I_7 & I & J \\
  M_7 & M_I & M_J
 \end{array} \right)  
 \left( \begin{array}{c c|c}
  T_7 & T_2 &  T \\
  M_{T_7} & M_{T_2}  & M_{T}
 \end{array} \right)
  \left( \begin{array}{c c|c}
  j_a & j_b &  I \\
  m_{j_a}  & m_{j_b}  & M_{I}
 \end{array} \right)   \left( \begin{array}{c c|c}
 \frac{1}{2}& \frac{1}{2} &  T_2 \\
  m_{t_a}  & m_{t_b}  & M_{T_2}
 \end{array} \right) \nonumber \\
&&   \sum_{\beta\gamma\delta}   \left( \frac{1}{2}\ \leftsub{\rm SD}{\left\langle ^7\textrm{Li} , \Omega_7 \left|\hat{a}_{\gamma}^{\dagger} \hat{a}_{\delta} ^{\dagger} \hat{a}_{a}  \hat{a}_{b} \hat{a}_{\beta}      \right| {}^8\textrm{Li} , \Omega'_8 \right\rangle}_{\rm SD} \left\langle   \beta a' \left| V \right| \gamma \delta \right\rangle  +  \leftsub{\rm SD}{\left\langle ^7\textrm{Li} , \Omega_7 \left|  \hat{a}^{\dagger}_{\beta}  \hat{a}^{\dagger}_{a'}    \hat{a}_{a}   \hat{a}_{\delta}  \hat{a}_{\gamma}   \right|  {}^8\textrm{Li} , \Omega'_8 \right\rangle}_{\rm SD}  \left\langle   \gamma  \delta  \left| V \right| \beta  b  \right\rangle  \right).
 \end{eqnarray}
 \end{widetext}

The main challenge in the 
computation of  Eqs.~(\ref{Summary_II}) and~(\ref{Summary_I_bis_total}) are the density matrix elements, which turn out to be time-consuming to calculate and cumbersome to store. %A way to handle the computational overhead was devised 
In Ref.~\cite{Navratil2011} we tackled this problem %we avoided their direct calculation 
by inserting a completeness relationship over $(A{-}5)$-body eigenstates %with the insertion of a resolution of the identity in the single-particle space 
between the triplet of creation operators and that of destruction operators in %the density matrix elements of 
Eq.~(\ref{V-kernel_5-compl}) and working with pre-computed coupled densities. %, which is derived making use of   the completeness of the $(A{-}5)$-body eigenstates. 
For systems with $A=6$ nucleons, this is a viable solution because the $(A-5)$-nucleon states are simply given by HO single particle states %$|n_{a} l_{a} j_{a} \frac{1}{2}\rangle$ 
and the reduced density matrix elements of Eq.~(\ref{V-kernel_5-compl}) involving $^4$He eigenstates are straightforward to calculate and store. However, systems with mass number bigger than 6 cannot be handled in the same way. Therefore, for the present work we implemented a new efficient `on the fly' calculation of the matrix elements of the three-body density of the target. % in a more efficient way, by calculating on the fly the three-body density matrices directly from the many-body wave functions in input, that is without the need to compute and store them in a separate calculation, which is not feasible for technical reason.
%The computation of the needed three-body density matrices from the input many-body wave function reduces in this way the number of the matrix densities effectively computed. 
This implementation relies on a hash algorithm, %implemented in the transition density code, 
which maps each configuration of a given NCSM target state in a unique sequence of bits of fixed size (typically an integer of 8 bytes for each species of nucleons). In this way the pairs and triplets of creation and annihilation operators in Eqs.~(\ref{Summary_II}) and (\ref{Summary_I_bis_total}) are implemented through bitwise operations, that allow to select efficiently the non-trivial density matrices for a given target state in input.

Finally, in the case of a NCSMC calculation besides the NCSM/RGM kernels one has to further compute overlap and Hamiltonian matrix elements between binary-cluster channel states and $A$-nucleon NCSM eigenstates of the composite nuclear system. For such matrix elements, which are comparatively much less computationally intensive, we will adopt the formalism and codes developed in Refs.~\cite{Baroni2013a,Baroni2013b}. For the sake of completeness, we outline the main features of the NCSMC approach in Appendix~\ref{App_1bis}.

\begin{table}[b]
\caption{\label{tab:tableNCSMstate}%
Ground-state and excitation energies of $^7$Li and $^8$Li calculated within the NCSM where the expansion of the nuclear wave function is truncated at $N_{\rm max}$= 6 and 8 in the HO basis and HO frequency $\hbar\Omega$=20 MeV, compared to the experiment. The values in 
the last column have been adjusted in order to reproduce the Q-value of the ${}^{7}$Li($d$,$p$)$^{8}$Li reaction, as explained in Section~\ref{sec:transfer d-Li7 results}.
% All the states of the $^8$Li included the ground state which decays $\beta^-$ to $^8$Be, are resonances and in the NCSM formalism are approximately described as eigenstates expanded in the HO basis. %
}
\begin{ruledtabular}
\begin{tabular}{c|c|dddd}
\textrm{Nucleus} &
\textrm{State} &
 \multicolumn{4}{c}{E (MeV)} \\
& \textrm{ J$^\pi$ } & \multicolumn{1}{c}{\textrm{$N_{\rm max}$= 6  }}&
\multicolumn{1}{c}{\textrm{$N_{\rm max}$= 8  }}&
\multicolumn{1}{c}{\textrm{Exp  }}&
\multicolumn{1}{c}{\textrm{Threshold  }}\\
& & & & &  \multicolumn{1}{c}{\textrm{${}^{7}$Li($d$,$p$)$^{8}$Li}} \\
\colrule
 $^7$Li & $\frac{3}{2}^-$ & -36.20 & -38.01 & -39.25  &-38.01\\
            & $\frac{1}{2}^-$ & -35.80 &  -37.64 & -38.77  & -37.53 \\
\colrule
$^8$Li & $2^+$ & -37.60 & -39.66&-41.28 & -40.04\\
            & $1^+$ & -36.36 & -38.47&-40.30 & -39.06\\
           & $3^+$ & -34.76 & -36.78 & -39.02 &  -37.78 \\
           & $0^+$\footnote{This state is a NCSM prediction not present in the experimental spectrum of $^8$Li} & -33.75 & -36.16 &   & -36.83\\
\end{tabular}
\end{ruledtabular}
\end{table}

\section{\label{sec:Results}Results} %for deuteron-$^7$L\MakeLowercase{i} reaction}
In this Section we apply the formalism developed in Refs.~\cite{Navratil2011,Navratil2012}, complemented with the NCSM/RGM kernels
as derived in Section~\ref{sec:formalism}, to the description of the $^9$Be spectrum above the $d$+$^7$Li threshold, the elastic scattering of deuterons on $^7$Li and protons on $^8$Li, %$\sim$17 MeV above the $^9$Be ground state.  
and the  $^7$Li$(d,p)^8$Li transfer reaction. %In particular we have performed two sets of calculations, corresponding to the elastic and transfer channels of the $d$-$^7$Li reaction, with the corresponding outgoing asymptotic states being $d$-$^7$Li and $p$-$^8$Li, respectively.
%Owing to the fact that we always work with a finite basis,  resonances in the compound nucleus 
%lying close to a breakup threshold can be faithfully described only in calculations %through explicit inclusion of 
%that include the corresponding mass partition in the model space. For instance, the resonances in the low-energy spectrum of $^9$Be are well described in the NCSM/RGM formalism when the basis is expanded in the $n$-$^8$Be binary channels~\cite{Langhammer2015}. Here we choose to expand the model space in the $d$-$^7$Li and $p$-$^8$Li cluster states, as we are interested in the description of the high-energy part of the spectrum, at $\sim$17 MeV above the $^9$Be ground state. 
Our choice for the interaction between nucleons is the chiral N$^3$LO NN potential of Ref.~\cite{Entem2003},
which is evolved through a similarity renormalization group (SRG) transformation with evolution parameter $\Lambda$=2.02 fm$^{-1}$. 
%With the same Hamiltonian we calculated energies and wave functions of the ground states and excited states of the projectile (the deuteron) and target states ($^7$Li and $^8$Li), which are obtained as linear combination of HO single-particle states in a NCSM variational calculation. 

Different from our earlier investigation of the low-energy spectrum of $^9$Be~\cite{Langhammer2015}, where the proximity to the $n$+$^8$Be breakup threshold justified a description based on expansions in  $n$-$^8$Be binary channels, %where we used $n$+$^8$Be binary channels to describe the low-energy spectrum of $^9$Be, 
here we are interested in excitation energies above $E_x=16.7$ MeV, where the $d-^7$Li channel opens immediately followed by the $p-^8$Li channel at $E_x=16.9$ MeV. For each $J^\pi T$ partial wave (we considered a maximum angular momentum of $J_{\rm max}=\frac{7}{2}$, for a total of 28 partial waves,  taking into account both positive and negative parities and the allowed values of the isospin $T$), the %NCSM/RGM study of the ${}^{7}$Li($d$,$p$)${}^{8}$Li deuteron stripping %to resonant states implies a reaction 
present study required complex coupled-channel calculations involving both the ($d,^7$Li) and ($p,^8$Li) mass partitions. 
%Truncating the partial-wave decomposition of the Schr\"odinger equation at $J_{\rm max}$=$\frac{7}{2}$, one has to deal with %This results in 
%28 different channels in the set of quantum numbers $J^\pi T$
%(14 for each parity, taking into account the allowed values of the isospin $T$).
%calculation where many channels are coupled: 
%Considering
Specifically, our model space included binary-cluster channels built upon 2 states ($\frac{3}{2}^-$ g.s. and $\frac{1}{2}^-$ first excited state) of the of $^7$Li and 4 states ($2^+$ g.s. and $1^+$, $3^+$, $0^+$ excited states) of the $^8$Li nuclei, as detailed in Table~\ref{tab:tableNCSMstate}.
For the deuteron we included the g.s.\ and described its non-resonant
continuum through the inclusion of discretized states, i.e. the pseudostates specified in Table~\ref{tab:d_pseudo}.
\begin{table}[t]%The best place to locate the table environment is directly after its first reference in text
\caption{\label{tab:d_pseudo}%
Ground-state and pseudostate energies of the deuteron calculated within the NCSM, with $N_{\rm max}= 8$ and 10 basis space, and HO frequency $\hbar\Omega$=20 MeV. In the calculations we included 4 pseudostate in the $^{3}S_1$-$^{3}D_1$ channel. }
\begin{ruledtabular}
\begin{tabular}{ldd}
\textrm {(Pseudo)state} &
\multicolumn{2}{c}{E (MeV)}\\
&  \multicolumn{1}{c}{\textrm{$N_{\rm max}$= 8  }}&
\multicolumn{1}{c}{\textrm{$N_{\rm max}$= 10  }}\\
\colrule
 g.s. & -1.96 & -2.12  \\
 1$^*$     & 9.91 &  8.36 \\
 2$^*$      & 15.22 &  12.82 \\
 3$^*$      & 33.24 &  26.6 \\
 4$^*$      & 40.20 &   33.23 \\                   
\end{tabular}
\end{ruledtabular}
\end{table}
As an example of the typical number of coupled channels we encountered, in the highest partial waves with $J=\frac{7}{2}$ %included in the present calculation, 
our model-space contained up to 60 binary channels specified by the collective index $\nu$ in Eqs.~(\ref{eq:trial},\ref{basis}). The size of the scattering matrix, which includes diagonal matrix elements describing elastic $d-^7$Li and $p-^8$Li scattering as well as off-diagonal matrix elements describing the $^7$Li$(d,p)^8$Li transfer reaction, is therefore considerable. %and we expect its off-diagonal terms of the matrix playing a decisive role in the description of the transfer process. In order
 
\begin{table}[b]%The best place to locate the table environment is directly after its first reference in text
\caption{\label{tab:NCSM_9Be}%
Ground-state and excited states energies of $^9$Be calculated within the NCSM, with truncation of the model-space at $N_{\rm max}$= 8 in the HO single-particle basis, and HO frequency $\hbar\Omega$=20 MeV. These states complement the cluster ones of the NCSM/RGM approach, producing the NCSMC model-space basis. Experimental values in parentheses correspond to the states with uncertain spin-parity assignment.}
\begin{ruledtabular}
\begin{tabular}{ldd|ldd}
\textrm {State} &
\multicolumn{2}{c}{E (MeV)} &
\textrm {State} &
\multicolumn{2}{c}{E (MeV)}\\
\multicolumn{1}{l}{\textrm{ J$^\pi$ }} &  \multicolumn{1}{c}{\textrm{Calc  }}&
\multicolumn{1}{c}{\textrm{Exp  }} &
\multicolumn{1}{l}{\textrm{ J$^\pi$ }} &  \multicolumn{1}{c}{\textrm{Calc  }}&
\multicolumn{1}{c}{\textrm{Exp  }}\\
\colrule
 $\frac{1}{2}^-$ & -53.82 & -55.38 & $\frac{1}{2}^+$ & -52.42 & -56.49 \\
     & -45.45 &   & &-42.69 &  \\
       & -41.57 &  -41.18 & & -40.29 &  \\
     $\frac{3}{2}^-$ & -57.45 & -58.16 &     $\frac{3}{2}^+$  & -48.80 & (-53.49) \\
     & -51.59 &  (-52.57) & & -43.25 &  \\
       & -46.13 &   & &-42.03 & \\
     & -41.59 &  -43.77 & & -37.57& \\   
          & -40.25 &   & &  &\\    
    $\frac{5}{2}^-$ & -54.78 & -55.73 & $\frac{5}{2}^+$ & -51.05 &  -55.11 \\
     & -49.49 &  (-50.22) & & -43.38 & \\
       & -44.05&  -46.35 & & -38.83 & (-41.49) \\
     & -43.15 &  (-44.37) & &-37.47 & \\   
          &  &  & &-37.23 & \\     
\end{tabular}
\end{ruledtabular}
\end{table}
Concerning the HO model space, we employed the frequency of $\hbar\Omega=20$ MeV and two truncations corresponding to a total number of excitations above the $2\hbar\Omega$ minimum-energy configuration of $N_{\rm max}=6$ and 8. To match the corresponding absolute number of HO  quanta, we described the deuteron in  $N_{\rm max}=8$, and 10 model spaces, respectively.
The values of the energies in Table~\ref{tab:tableNCSMstate} and those of the g.s. energy of the deuteron
in Table~\ref{tab:d_pseudo} show that the calculation with the largest model-space basis corresponding
to $N_{\rm max}= 8$, is not converged. This sets the computational limit for the present application
of our approach. %to the transfer reactions within the present implementation.

Finally, to explore the interplay between the $d-^7$Li and $p-^8$Li channels and the effect of short-range 9-body correlations that are not efficiently included in the 
cluster wave functions of the NCSM/RGM approach, for the case of the elastic $d-^7$Li and $p-^8$Li scattering we also performed  `uncoupled' calculations (in the first case within the ($d,^7$Li) and in the second case within the ($p,^8$Li) mass partitions) with and without the inclusion of $^9$Be eigenstates computed within the NCSM approach.  The number of $^9$Be negative-parity states used for this calculation are 30 (12) for $N_{\rm max}$= 6 (8) and for the positive parity we added 21 (12) for $N_{\rm max}$= 6 (8). States and corresponding energies for $N_{\rm max}$= 8 are specified in Table~\ref{tab:NCSM_9Be}. %With this application of the NCSMC method~\cite{Baroni2013a,Baroni2013b}, we aim at capturing the short-range many-body physics of the $A$=9 nucleon system, which is not efficiently included in the cluster wave functions of the NCSM/RGM approach. For the sake of completeness, we outline the main features of the NCSMC approach in Appendix~\ref{App_1bis}.

We start the discussion of our results by analyzing the scattering eigenphase shifts obtained in the coupled NCSM/RGM calculation in Sec.~\ref{sec:eigenphaseshifts}. Next, we discuss the elastic $p-^8$Li and $d-^7$Li scattering in Secs.\ref{sec:elastic p-Li8 results} and ~\ref{sec:elastic d-Li7 results}, respectively. Finally, the computed $^7$Li$(d,d)^7$Li  and $^7$Li$(d,p)^8$Li cross sections are presented in Secs.~\ref{sec:elastic d-Li7 cross section} and  \ref{sec:transfer d-Li7 results}, respectively.

\subsection{\label{sec:eigenphaseshifts}Eigenphase shifts}
\begin{figure}[t]
{\includegraphics[width=\columnwidth,keepaspectratio]{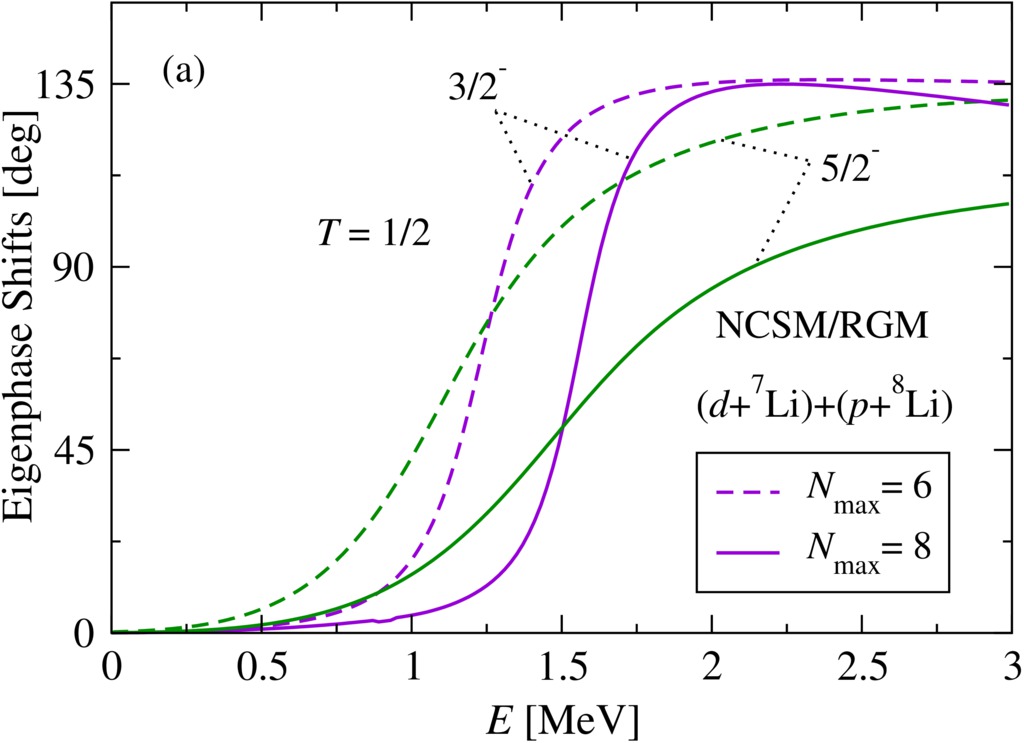}}% Here is how to import EPS art
\quad
{\includegraphics[width=\columnwidth,keepaspectratio]{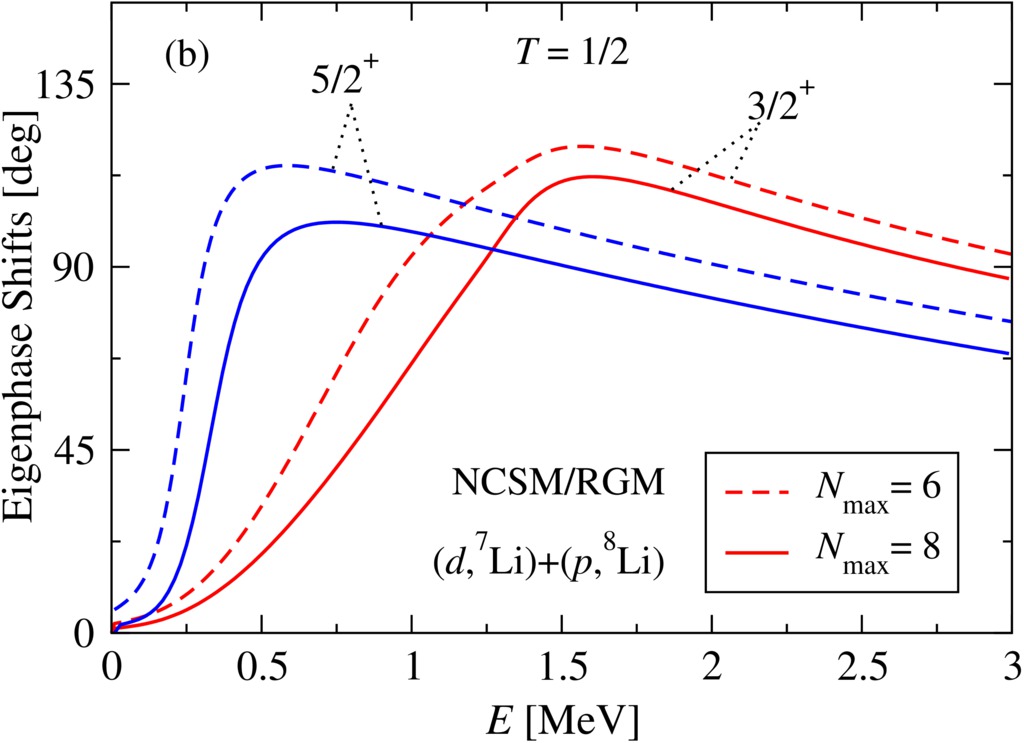}}
\caption{\label{eigen_minus_plus}Calculated (a) negative and (b) positive-parity eigenphase shifts within the coupled ($d,^7$Li)+($p,^8$Li) NCSM/RGM basis, as a function of the relative kinetic energy in the c.m.\ frame with respect to the $p$+$^8$Li threshold. The SRG-N$^3$LO NN potential with $\Lambda$=2.02 fm$^{-1}$, and the HO frequency of $\hbar\Omega$=20 MeV were used.}
\end{figure}
\begin{figure}[t]
\includegraphics[width=\columnwidth,keepaspectratio]{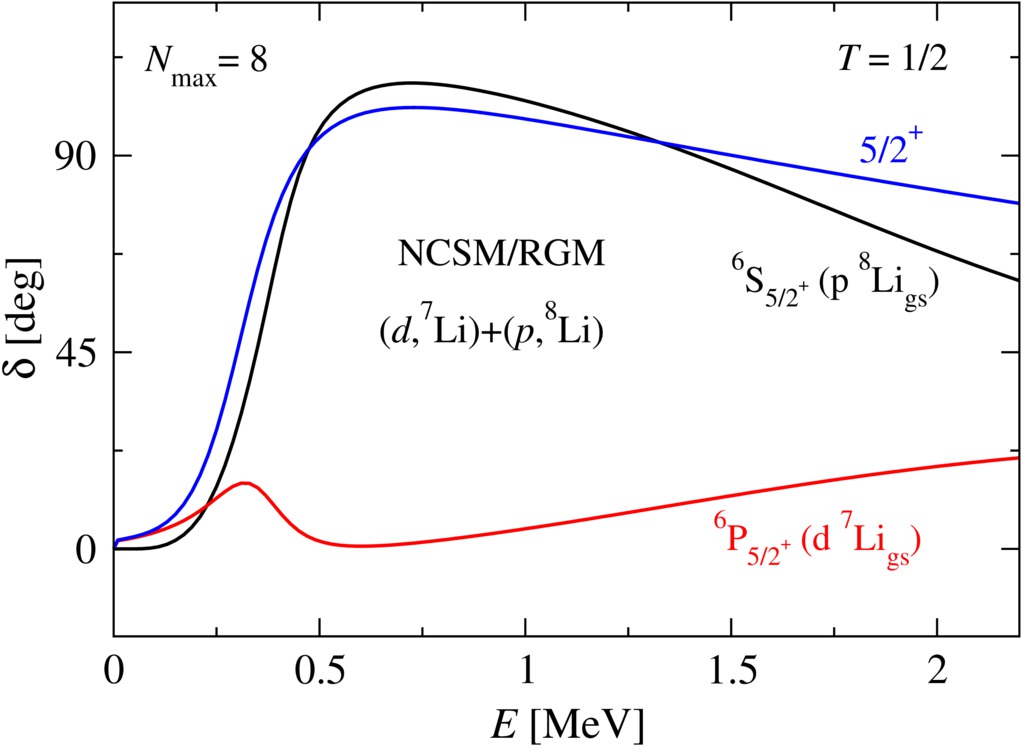}%
\caption{\label{phase_shift_5half} Eigenphase shifts for the $J^\pi T = \frac{5}{2}^+\, \frac12$ partial wave  (solid blue line) compared to the $d$-$^7$Li (solid red line) and $p$-$^8$Li (solid black line) elastic phase shifts contributing to the same partial wave through a $P$- and $S$-wave respectively. All results were obtained  within the coupled ($d,^7$Li)+($p,^8$Li) NCSM/RGM basis, and are plotted as a function of the relative kinetic energy in the c.m.\ frame with respect to the $p$+$^8$Li threshold.}
\end{figure}

%
%To determine the dominant partial waves involved in the $^7$Li$(d,p)^8$Li transfer reaction,  we first analize 
The scattering eigenphase shifts convey information about the scattering matrix as a whole. In the present  coupled calculations, they encompass information about the  $d-^7$Li and $p-^8$Li elastic scattering as well as the $^7$Li$(d,p)^8$Li transfer reaction. %In this way we determine which are the dominant partial waves involved in the reaction.
A selection of our computed eigenphase shifts for negative- and 
positive-parity states is presented in Fig.~\ref{eigen_minus_plus}(a) and (b), respectively. For clarity of the figure we only show the $T=\frac{1}{2}$ resonant eigenphase shifts for $J=\frac{3}{2}$ and $\frac{5}{2}$, which are the main responsible for  the strength of the peaks in the $^7$Li$(d,p)^8$Li cross section.   Even though the curves at $N_{\rm max}$= 8 are not converged, the comparison between the eigenphase shifts at $N_{\rm max}$= 6 and 8 shows the preliminary trend of the convergence. Near the $p$+$^8$Li threshold (corresponding to $0$ energy in the figure), the dominant eigenphase shift appears in the $\frac{5}{2}^+$ partial wave. Figure~\ref{phase_shift_5half} further compares the two main phase shifts contributing to this latter resonant state.  
%\begin{figure}[b]
%\includegraphics[width=\columnwidth,keepaspectratio]{/Users/francescoraimondi/DEUTERON_TRANSFER_PAPER/PLOTS/EPS_FIG/eigenphase_shift_p8Li_d7Li_srg-n3lo2.02_20_Nmax_6_8_4st_2st_RGM_MAIN_smooth_PLUS2.eps}%
%\caption{\label{eigen_plus} A figure caption. The figure captions are
%automatically numbered.}
%\end{figure}
%
%\begin{figure}[b]
%\includegraphics[width=\columnwidth,keepaspectratio]{/Users/francescoraimondi/DEUTERON_TRANSFER_PAPER/PLOTS/EPS_FIG/eigenphase_shift_p8Li_d7Li_srg-n3lo2.02_20_Nmax_6_8_4st_2st_RGM_MAIN_smooth_MINUS2.eps}%
%\caption{\label{eigen_minus} A figure caption. The figure captions are
%automatically numbered.}
%\end{figure}
%In order to assess the importance of the relevant channels in the low-energy dominant $\frac{5}{2}^+$ partial wave,
%we show in Fig.~\ref{phase_shift_5half} the corresponding resonant phase shifts. 
These are due to the %two mass partition channels
$d$+$^7$Li and $p$+$^8$Li  mass partitions in relative $P$- and $S$-wave motion, respectively, with the $S$-wave coupling of the proton with the g.s. of the $^8$Li having a clear resonant behavior.

In the following we discuss elastic and transfer processes separately.

\subsection{\label{sec:elastic p-Li8 results}Elastic $p-^8$Li scattering phase shifts}

\begin{figure}
{\includegraphics[width=\columnwidth,keepaspectratio]{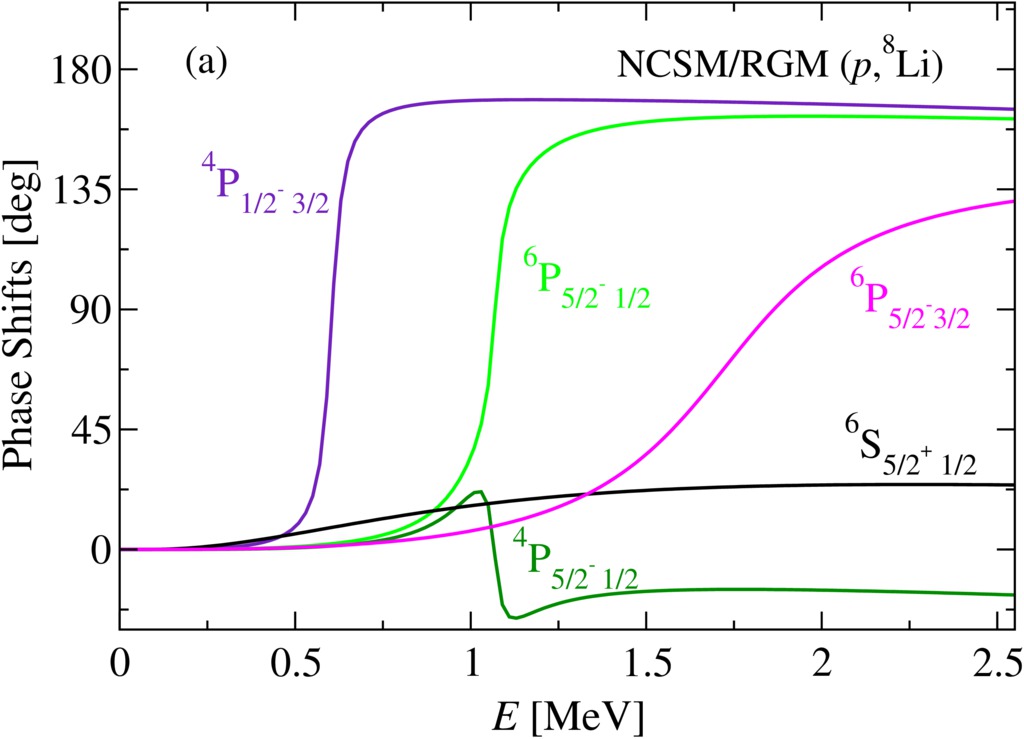}}
\quad
{\includegraphics[width=\columnwidth,keepaspectratio]{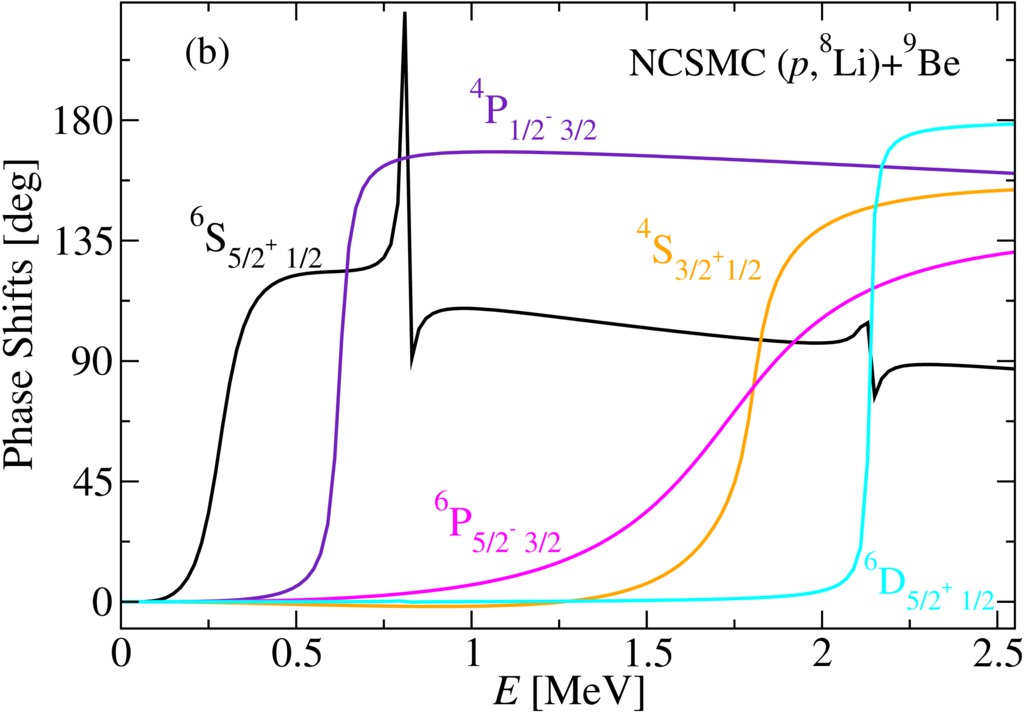}}
\caption{\label{phase_shift_p_Li8} Calculated $p-^8$Li elastic phase shifts within the (a) ($p,^8$Li) NCSM/RGM and (b) ($p,^8$Li)+$^9$Be NCSMC  bases as a function of the relative kinetic energy in the c.m.\ frame. The SRG-N$^3$LO NN potential with $\Lambda=2.02$ fm$^{-1}$, the $N_{\rm max}= 8$ basis size, and the HO frequency of $\hbar\Omega=20$ MeV were used. The $^9$Be NCSM eigenstates used in the NCSMC calculation are specified in Table~\ref{tab:NCSM_9Be}.}
\end{figure}
It is instructive to compare the $p$-$^8$Li $^6S_{\frac{5}{2}^+}$ phase shifts of Fig.~\ref{phase_shift_5half} with those resulting from calculations without the $d$-$^7$Li channels, shown in Fig.~\ref{phase_shift_p_Li8}(a). In the absence of coupling to the $d$-$^7$Li mass partition the $S$-wave phase shifts are strongly suppressed. %quenched and show at most a weak resonant behavior, with  
The elastic $p$-$^8$Li scattering below 1 MeV is instead
dominated by the $\frac{1}{2}^-$ partial wave with isospin $T=\frac{3}{2}$, which is associated with the $^9$Be resonant state at 16.98 MeV above the g.s.\ energy. Another important resonance with isospin $T=\frac{3}{2}$ reproduced in our calculation belongs to the $\frac{5}{2}^-$ partial wave and is related to the 18.65 MeV resonance in $^9$Be, the structure of which has been studied with proton scattering at 180 MeV~\cite{Dixit1991}.
We note that the isospin $T=\frac{3}{2}$ channel does not contribute to the transfer reaction process, owing to the fact that the $d$-$^7$Li  mass partition can only be coupled to isospin $T=\frac{1}{2}$.
Fig.~\ref{phase_shift_p_Li8}(a) also shows two resonant phase shifts at $\sim1$ MeV in the  $J^\pi T=\frac{5}{2}^-\,\frac{1}{2}$ partial wave. However, as for the $^6S_{\frac{5}{2}^+}$ phase shifts before, a purely ($p,^8$Li) NCSM/RGM calculation does not provide a complete picture for this partial wave. 

%\begin{figure}[b]
%\includegraphics[width=\columnwidth,keepaspectratio]{PLOTS/EPS_FIG/phase_shift_1H8Li_1H8Li_srg-n3lo2p02_20_8_4st_4st_RGM_MAIN2.eps}
%\caption{\label{phase_shift_p_Li8} Calculated NCSM/RGM $p$-$^8$Li elastic phase shifts as a function of the relative kinetic energy of the proton-$^8$Li system in the c.m. frame. The SRG-N$^3$LO NN potential with $\Lambda$=2.02 fm$^{-1}$, the $N_{\rm max}$= 8 basis size, and the HO frequency of $\hbar\Omega$=20 MeV were used.}
%\end{figure}
The $p$-$^8$Li elastic phase shifts %The relative importance of the opposite parities in the $J$=$\frac{5}{2}$ partial wave of the $p$-$^8$Li channel, changes also when 
are also influenced by short-range many-body correlations. This can be observed by comparing the results of Fig.~\ref{phase_shift_p_Li8}(a) with those of Fig.~\ref{phase_shift_p_Li8}(b), obtained in a NCSMC model space spanned by the same set of $p$-$^8$Li channel states and the $^9$Be eigenstates of Table~\ref{tab:NCSM_9Be}.  We see once again an enhancement of the $^6S_{\frac{5}{2}^+}$ phase shifts, which becomes the first strong resonance above the proton-$^8$Li threshold. It should be noted that this behavior is not in contradiction with the trend exhibited by the coupled NSCM/RGM calculation of Fig.~\ref{phase_shift_5half}.  Indeed, the inclusion of $\frac{5}{2}^+$ NCSM eigenstates of the $^9$Be nucleus partly makes up for the missing ($d,^7$Li) mass partition, albeit failing to describe the portion of resonance escape width due to this decay channel. The same argument also applies to all other $T=\frac{1}{2}$ phase shifts in which the ($d,^7$Li) mass partition plays a role such as the other two positive-parity resonances with $J=\frac{3}{2}$ and $\frac{5}{2}$ appearing at energies above 1.5 MeV, in the $S$- and $D$-waves respectively. Furthermore, in the NCSMC calculation the $\frac{5}{2}^-$ state becomes bound with respect to the $p+^8$Li threshold, whereas the resonances with isospin $T=\frac{3}{2}$ are left unchanged, owing to the fact that we did not add $^9$Be NCSM eigenstates in that isospin channel.

In summary, the analysis of this section indicates that the $T=\frac12$ $p$-$^8$Li elastic phase shifts are strongly influenced by the coupling to $^9$Be eigenstates. For the $^6S_{\frac{5}{2}^+}$ partial wave the observed enhancement of the low-lying $p$-$^8$Li resonance is an effect of the coupling with the $d-^7$Li decay channel, and can be described well within the coupled NCSM/RGM calculation of Fig.~\ref{phase_shift_5half}. A more complete calculation including also higher-energy decay modes would be required for the resonances found above $\sim1.5$ MeV. However, as it will be clear when we discuss the computed total ${}^{7}$Li($d$,$p$)${}^{8}$Li cross section, such resonances do not contribute in the peak region we are primarily interested in, that is around 17.30 MeV above $^9$Be ground state.

\subsection{\label{sec:elastic d-Li7 results}Elastic $d-^7$Li scattering phase shifts}
\begin{figure}[t]
\includegraphics[width=\columnwidth,keepaspectratio]{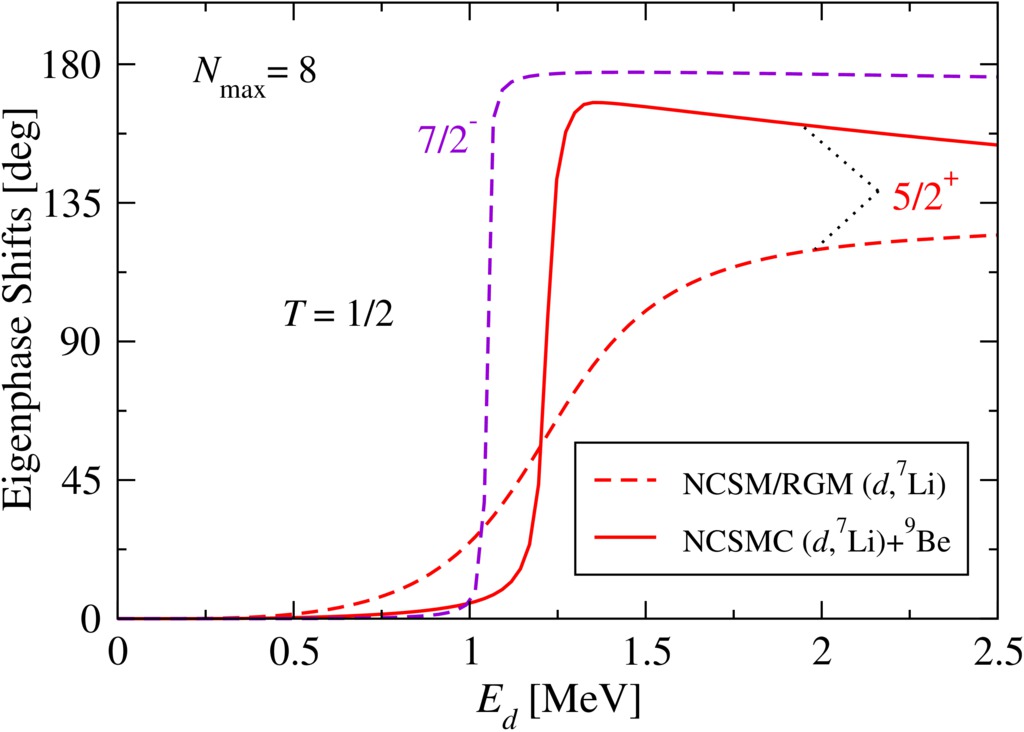}%
\caption{\label{eigen_5plus_7minus_d7Li} Calculated $d-^7$Li eigenphase shifts in the $J^\pi T= \frac{5}{2}^+ \, \frac12$ and $\frac{7}{2}^-\, \frac12$ partial waves as function of the kinetic energy of the deuteron projectile in the laboratory system, resulting from ($d,^7$Li) NCSM/RGM (dashed lines) and ($d,^7$Li)+$^9$Be NCSMC (solid line) calculations. The SRG-N$^3$LO NN potential with $\Lambda=2.02$ fm$^{-1}$, the $N_{\rm max}= 8$ basis size, and the HO frequency of $\hbar\Omega=20$ MeV were used. For the $\frac{7}{2}^-$ channel the NCSMC and NCSM/RGM eigenphase shifts are identical because the adopted set of $^9$Be eigenstates of  Table~\ref{tab:NCSM_9Be} did not include states for this partial wave. }
\end{figure}

\begin{figure}[t]
\includegraphics[width=\columnwidth,keepaspectratio]{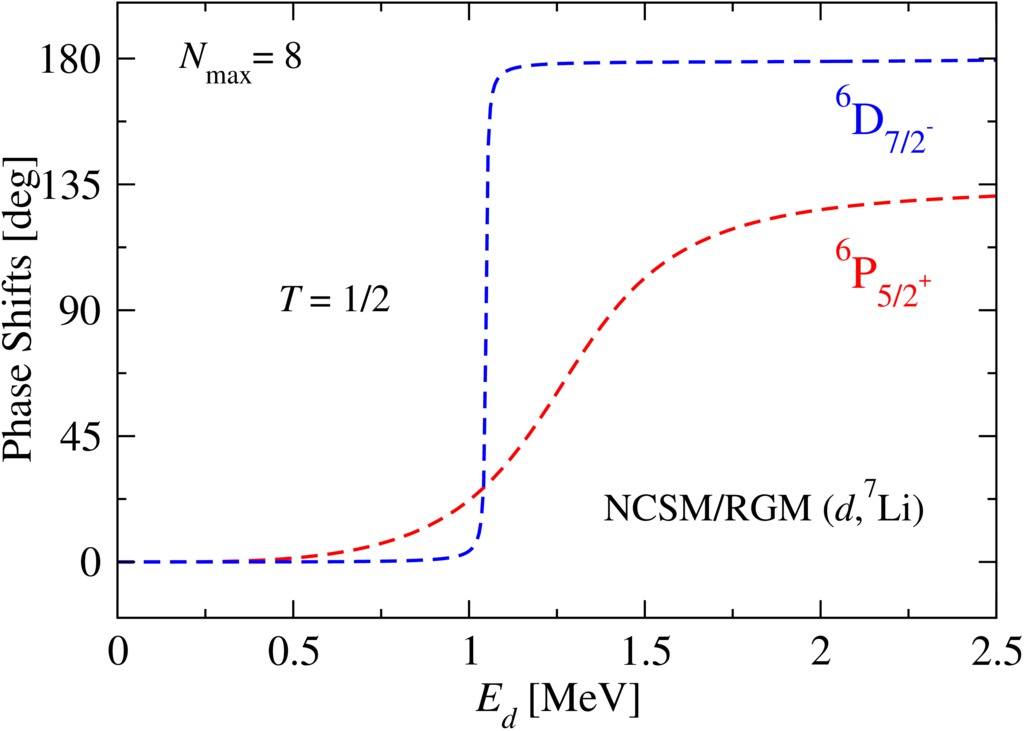}%
\caption{\label{phase_5plus_d7Li} Resonant $d-^7$Li $^6D_{\frac{7}{2}}$ and $^6P_{\frac{5}{2}}$ phase shifts %in the $J^\pi$= $\frac{5}{2}^+$ and $\frac{7}{2}^-$ partial waves 
as functions of the kinetic energy of the deuteron projectile in the laboratory frame, calculated within the ($d,^7$Li) NCSM/RGM basis.
The SRG-N$^3$LO NN potential with $\Lambda$=2.02 fm$^{-1}$, the $N_{\rm max}$= 8 basis size, and the HO frequency of $\hbar\Omega$=20 MeV were used.}
\end{figure}

%Before computing the transfer reaction, we study 
%We start our discussion with a study of 
Working within a ($d,^7$Li) NCSM/RGM model space we find that, above the $d$+$^7$Li threshold and below the deuteron
breakup energy the $^9$Be spectrum presents two resonances,  one each in the $J^\pi$= $\frac{5}{2}^+$ and $\frac{7}{2}^-$  partial waves. The corresponding eigenphase shifts are plotted in Fig.~\ref{eigen_5plus_7minus_d7Li} as functions of the kinetic energy
of the deuteron in the laboratory frame. 
This picture is corroborated by the results (also shown in Fig.~\ref{eigen_5plus_7minus_d7Li}) obtained in calculations carried out in a NCSMC  model space additionally incorporating the $^9$Be NCSM eigenstates of Table~\ref{tab:NCSM_9Be}. For the $\frac{7}{2}^-$ channel the two calculations produce identical eigenphase shifts owing to the absence of  $^9$Be states in the NCSM portion of the basis. At the high excitation energies considered in this work, the considerably large density of $^9$Be levels made it extremely difficult to identify and extract all relevant partial waves.  %In Fig.~\ref{eigen_5plus_7minus_d7Li} we plot the attractive eigenphase shifts  as function of the kinetic energy
%of the deuteron in the laboratory frame for the two partial waves $J^\pi$= $\frac{5}{2}^+$ and $\frac{7}{2}^-$ in which we find low-lying resonances. In particular we focus on the $J^\pi$= $\frac{5}{2}%^+$ and $\frac{7}{2}^-$ partial waves, corresponding to the resonant states of $^9$Be above $d$+$^7$Li threshold and below the deuteron
%breakup energy. 
For the $\frac{5}{2}^+$ eigenphase shifts, the short-range correlations introduced in the nuclear wave function through the $^9$Be NCSM eigenstates leave the position of the resonance unchanged but lead to a much narrower  width, %the excitation is turned into a very narrow resonance, 
as shown by the steep NCSMC curve. While this difference points to a somewhat slow convergence of the NCSM/RGM calculation, it is also important to note that  without the explicit inclusion of the (nearby) $p$+$^8$Li particle-decay channel the width of the $\frac{5}{2}^+$ resonance is artificially underestimated in the NCSMC.
Indeed, the coupling to the $p$-$^8$Li mass partition has an opposite effect on the  the ${\frac{5}{2}^+}$  $d$-$^7$Li elastic phase shifts, that is a quenching of the resonance. This can be observed by comparing the coupled NSCM/RGM calculation of Fig.~\ref{phase_shift_5half} with the $^6P_{\frac{5}{2}}$ elastic phase shifts of Fig.~\ref{phase_5plus_d7Li}, which do not include the effect of the $p+^8$Li channel.

%In general, the long lifetime of the resonances in the elastic calculation, which misses the decay channel in the $p$+$^8$Li mass partition, can be an artificial effect of the restricted model-space basis.

%                              
Upon further analysis, %of the most important channels for the $^7$Li($d$,$d$)$^7$Li reaction, 
we found that  the $\frac{5}{2}^+$ and $\frac{7}{2}^-$ scattering states are dominated by $d$-$^7$Li channels with the relative motion respectively in $D$ and $P$ wave. This becomes evident when comparing the NCSM/RGM results of Fig.~\ref{eigen_5plus_7minus_d7Li} with those of Fig.~\ref{phase_5plus_d7Li}, showing the $^6D_{\frac{7}{2}}$ and $^6P_{\frac{5}{2}}$ diagonal phase shifts.
% we show in 
%Fig.~\ref{phase_5plus_d7Li} the resonant phase shifts for the $\frac{5}{2}^+$ and $\frac{7}{2}^-$ partial waves. In particular the $^6D_{\frac{7}{2}}$ phase shift show a clear resonant behavior that %plays a significant role 
%in the excitation function of the reaction, as it will be clear in the following discussion of the differential cross section.
%
%\begin{figure}[t]
%\includegraphics[width=\columnwidth,keepaspectratio]{PLOTS/EPS_FIG/dsigma_dOmega_90deg_RGM_NCSMC_dLi7_Be9_Nmax8_PLOT2.eps}%
%\caption{\label{dsigma_d7Li} Elastic $d$-$^7$Li differential cross section in the c.m. frame as function of the
%kinetic energy of deuterons in the laboratory system, compared to the experimental data of Ref.~\cite{Ford1964}. The results in the NCSMC calculation (solid line) are obtained by enlarging the model-space basis of the NCSM/RGM calculation (dashed line) 
%with the inclusion of 12 negative and 12 positive parity $^9$Be states obtained from a NCSM calculation.}
%\end{figure}

\subsection{\label{sec:elastic d-Li7 cross section}$^7$Li$(d,d)^7$Li cross section} 
The $^7$Li$(d,d)^7$Li cross section below the deuteron breakup energy %. This process 
has been measured with the aim to investigate the resonant states of $^9$Be above the $d$+$^7$Li threshold~\cite{Ford1964,Imhof1965,Lombaard1974}.
Here, we will %present three sets of 
compare the differential cross section at the deuteron c.m.\ scattering angle of $90^\circ$ of Ford~\cite{Ford1964}  with the results of calculations performed within a model space spanned exclusively by $d-^7$Li channel states, as well as 
with those obtained by further including either $^9$Be eigenstates or $p-^8$Li channel states.   

\begin{figure}[t]
\includegraphics[width=\columnwidth,keepaspectratio]{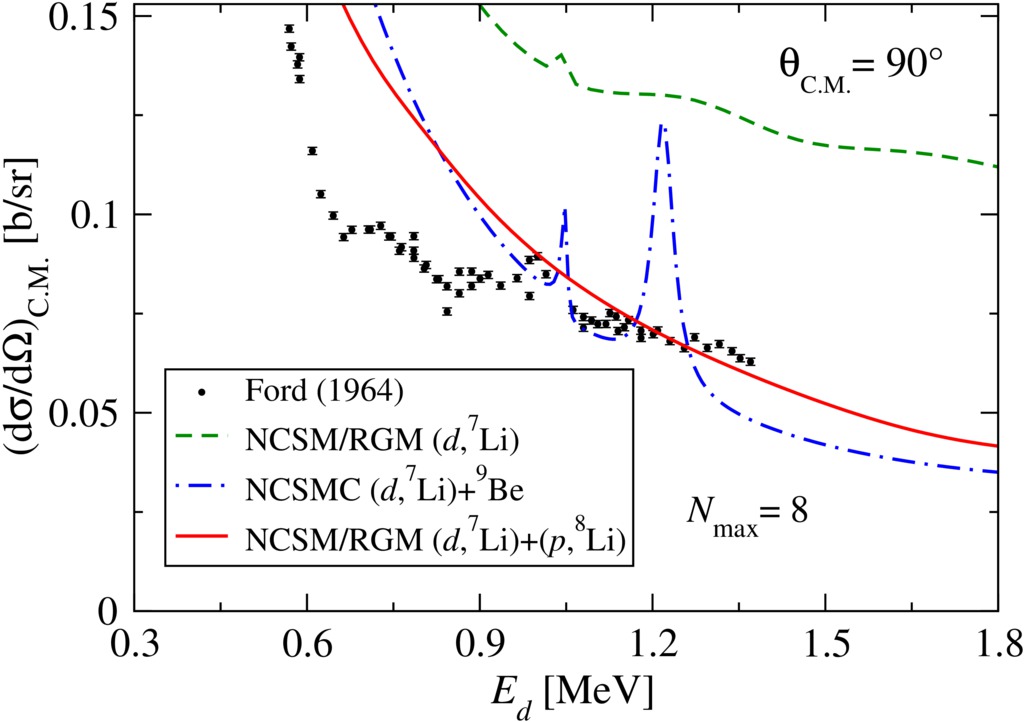}%
\caption{\label{dsigma_d7Li} Computed $^7$Li$(d,d)^7$Li differential cross sections in the c.m.\ frame at the deuteron scattering angle of $90^\circ$ as function of the
kinetic energy of deuterons in the laboratory system, compared to the experimental data of Ref.~\cite{Ford1964}. The three sets of theoretical curves correspond to calculations within the ($d,^7$Li) NCSM/RGM (green dashed line), ($d,^7$Li)+$^9$Be NCSMC (blue dash-dotted line), and ($d,^7$Li)+($p,^8$Li) NCSM/RGM (red solid line) model spaces.}
\end{figure}

The resonant behavior in the $^6D_{\frac{7}{2}}$ and $^6P_{\frac{5}{2}}$ phase shifts of Fig.~\ref{eigen_5plus_7minus_d7Li} %and \ref{phase_5plus_d7Li} 
explains the 
two peaks at around
1 and 1.2 MeV, respectively, observed in the
%these two resonances are responsible for the 
%two-peak structure of the %($d,^7$Li) NCSM/RGM and 
NCSMC differential cross sections shown in Fig.~\ref{dsigma_d7Li} (blue dash-dotted line). 
%This result was obtained by truncating the partial wave decomposition of the Schr\"odinger equation at $J_{\rm max}$=$\frac{7}{2}$. 
Compared to the ($d,^7$Li) NCSM/RGM results (green dashed line), the first peak is roughly the same (save for differences in the energy grids used in the two calculations) while the second becomes much more pronounced and narrower %is  in the In particular, the second peak is not present in the %compared to the 
%($d,^7$Li) NCSM/RGM results (green dashed line), there 
due to the inclusion of the  $^9$Be eigenstates, which have also the effect of  %leads to
%In particular, the %presence of 
%two peaks at around
%1 and 1.2 MeV due to the $J^\pi$=$\frac{7}{2}^-$ and $\frac{5}{2}^+$ partial waves, respectively,
%  is clearly visible in the second, where the inclusion of the  $^9$Be eigenstates 
%  and brings 
bringing the calculated differential cross section closer in magnitude to the experimental data of Ford~\cite{Ford1964}. In this experiment, the %elastic scattering of 
$^7$Li($d$,$d$)$^7$Li cross section %deuterons with laboratory energy in the $0.4-1.8$ MeV range 
shows an enhancement %of the cross section 
at about 0.8 %and 1 
MeV  and a resonance around 1 MeV, which were found to be compatible with deuterons traveling in %a fit based on the assumption of 
$S$- %wave %deuterons for the first 
and $P$-wave, %deuterons, 
respectively. 
 %for the second~\cite{Ford1964}. 
%The two peaks in the calculated NCSMC differential cross section are due to the $J^\pi$=$\frac{7}{2}^-$ and $\frac{5}{2}^+$ partial waves respectively. 
At the same time, the coupling to the $p-^8$Li channel has also a significant impact on the $^7$Li$(d,d)^7$Li cross section of Fig.~\ref{dsigma_d7Li}, where the solid red (dashed green) line represents the %differential cross section at 90 degree of the $d$-$^7$Li scattering 
NCSM/RGM result obtained with (without) the ($p,^8$Li) mass partition. Specifically, such coupling has the effect  of moving down the computed NCSM/RGM curve, bringing it in fairly good agreement with the experimental data in the region above 1 MeV, while the $\frac{5}{2}^+$ peak observed
in the NCSMC differential cross section is not present owing to the quenching of the $^6P_{\frac{5}{2}}$ resonance. 

Reconciling these experimental and theoretical points of view %to arrive at an interpretation of these resonant peaks 
is not easy. On one hand, the inclusion of NCSM $^9$Be energy eigenstates in the NCSMC calculation enhances the impact of the short-range correlations difficult to describe in terms of binary-cluster basis states. On the other hand,  the lifetime of the resonances is artificially increased by the lack of $p$-$^8$Li cluster states, which would provide a 
channel of decay lying just above the $d$-$^7$Li threshold.  Therefore, while we currently are not in the position of performing a more conclusive NCSMC study including also $p$-$^8$Li channels, we can tentatively associate the first calculated peak (corresponding to a $\frac{7}{2}^-$ state) to the experimental enhancement of the cross section  around 0.8 MeV and the second one (corresponding to a $\frac{5}{2}^+$ state) to the experimental resonance at 1 MeV. This interpretation would imply that the computed cross section is shifted to higher energies and the two peaks are narrower and further apart from each other than in experiment. At the same time, 
the relative importance of the $p-^8$Li $S$-wave channel over the $d-^7$Li $P$-wave one in the dominant $\frac{5}{2}^+$ partial wave observed in the coupled NCSM/RGM calculation could explain why the resonant structure of $^9$Be is hardly visible in the experimental ${}^{7}$Li($d$,$d$)${}^{7}$Li elastic data (see experimental points in Figs.~\ref{dsigma_d7Li}), whereas it is clearly pronounced in the transfer process, as it will be clear from the  discussion in the following section.

The fact that the microscopic Hamiltonian in our present calculation is incomplete, i.e.\ that we do not include 3N forces, may in part be at the origin of the disagreement between computed and measured elastic cross sections % with the measured one 
observed in Fig.~\ref{dsigma_d7Li} also in the case of the more complete NCSMC model space. Indeed, already the computed NCSMC $^9$Be g.s.\ %energy of the $^9$Be 
is found at $-17.4$ MeV (with respect to the $d+^7$Li threshold),  overbound by 4$\%$ with respect to the experimental value. It is well known~\cite{Jurgenson2009,Jurgenson2011,Roth2011} that the lack of higher-body terms in the microscopic Hamiltonian leads to a dependence of computed observables on the SRG flow parameter. In this respect a heuristic choice of the flow parameter  $\Lambda$ should be guided by the strategy of minimizing the impact of the bare 3N forces through the onset of higher-body terms induced by the SRG evolution of the NN interaction.  This can work provided that the interplay between bare and induced forces goes in the direction of a mutual cancellation, which is not \textit{a priori} guaranteed.  Our present choice of the flow parameter ($\Lambda$=2.02 fm$^{-1}$) is motivated by the study of the dependence of $^4$He binding energy on $\Lambda$~\cite{Jurgenson2009},
but it appears not to be the optimal one in order to minimize the impact of the missing 3N forces in in the present case. %of p-shell nuclei. 

\subsection{\label{sec:transfer d-Li7 results}${}^{7}$Li($d$,$p$)${}^{8}$Li transfer reaction }

%To understand the role played by the off-diagonal elements of the scattering matrix  

%\begin{figure}[b]
%\includegraphics[width=\columnwidth,keepaspectratio]{PLOTS/EPS_FIG/phase_shift_p8Li_9Be_srg-n3lo2p02_20_8_4st_NCSMC_SELECTED2.eps}
%\caption{\label{phase_shift_p_Li8_NCSMC} As in Fig.~\ref{phase_shift_p_Li8} for the NCSMC calculation.}
%\end{figure}

\begin{figure}[t]
\includegraphics[width=\columnwidth,keepaspectratio]{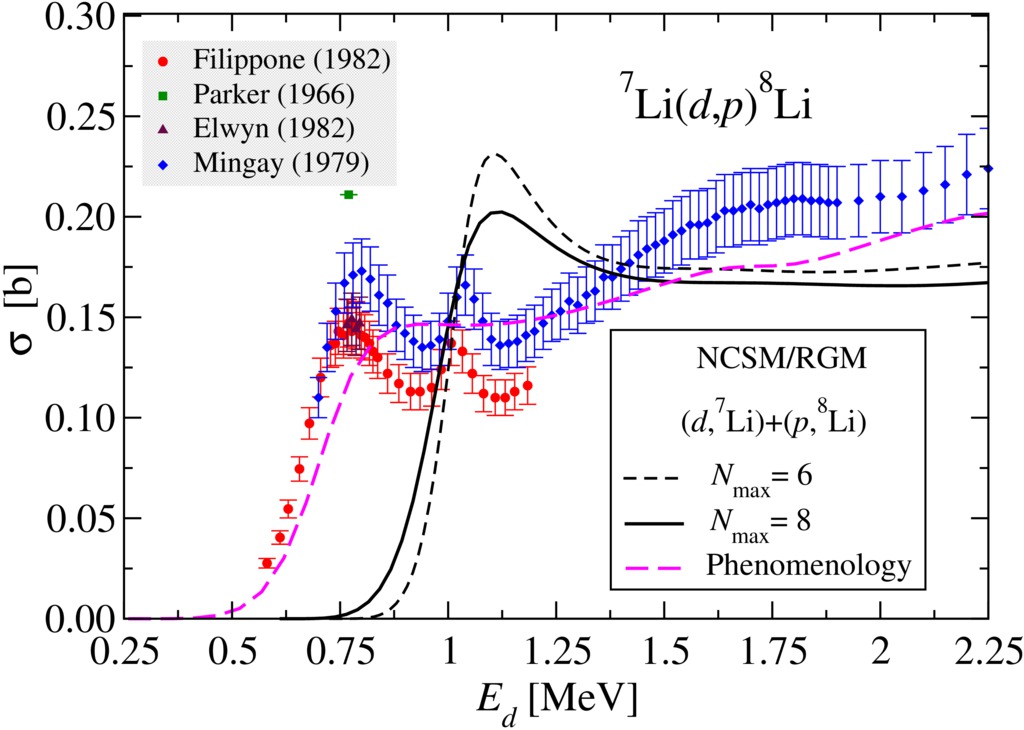}%
\caption{\label{sigma_tot} $^7$Li($d$,$p$)$^8$Li integrated cross section for deuteron laboratory energies up to 2.25 MeV computed within the NCSM/RGM approach at $N_{\rm max}= 6$ (thin-dashed line) and 8 (solid line) compared to the experimental data from Refs.~\cite{Parker1966,Mingay1979, Elwyn1982,Filippone1982} (symbols). Also shown as a thick dashed magenta line is the result of the NCSM/RGM phenomenology approach (see text for details).}
\end{figure}

In Fig.~\ref{sigma_tot}, we compare our calculated ${}^{7}$Li($d$,$p$)${}^{8}$Li total cross section to the experimental data of Refs.~\cite{Parker1966,Mingay1979, Elwyn1982,Filippone1982} for deuteron energies in the laboratory frame up to about 2.3 MeV. 
We include approximately breakup effects for the deuteron with pseudostates displayed in Table~\ref{tab:d_pseudo}, and we consider only the low-energy
part of the excitation function, i.e. below the breakup threshold of the deuteron. Moreover, for higher energies of the projectile we can expect a bigger impact of other channels which are missing in our present calculation, such as 
neutron-$^8$Be and triton-$^6$Li.

%The disagreement with the experimental cross section data may be due to the non-convergence in the HO single-particle basis used to expand the nuclear many-body wave function, given that
%our calculation with the largest model-space basis corresponds to $N_{\rm max}$= 8. 

In comparing the cross section to the experimental data, it should be kept in mind that at the largest feasible model space $N_{\rm max}= 8$ our many-body wave function has likely not reached convergence yet with respect to the size of %single-particle 
the HO basis. Going from $N_{\rm max} = 6$ to 8, the height of the first peak
of the calculated cross section moves towards the experimental recommended value of 0.147$\pm$0.011 b~\cite{Adelberger1998}. This peak, found in nature at the deuteron kinetic energy of 0.78 MeV as a broad
structure of width $\Gamma\approx$ 0.2 MeV, is used to determine the mean areal density of $^7$Be atoms in the targets used in experimental studies of the $^7$Be($p$,$\gamma$)$^8$B radiative
capture~\cite{Adelberger1998}. The spin-parity assignment of this important resonant state is experimentally uncertain:
Phenomenological R-matrix analyses~\cite{Decharge1972,Friedland1971} based on the measurement of the angular distributions of $\alpha$ particles from the $\beta$ decay of the $^8$Li produced in the $d-^7$Li transfer reaction, are compatible with either a $J=\frac{3}{2}$ or a $\frac{5}{2}$ assignment for the total spin, but disagree on the determination of the intrinsic parity of the state. Our calculation supports a $\frac{5}{2}^+$ spin-parity assignment, as illustrated in the phase shifts plot of Fig.~\ref{phase_shift_5half}, suggesting a reaction mechanism dominated by the coupling of the $P$-wave $d-^7$Li incoming channel to the $S$-wave in the $p-^8$Li exit channel.

Our calculation overestimates the position of the first peak by about 0.33 MeV. This is in line with %The truncation in the model-space basis is the possible reason of 
the overestimation of both Q-value
and threshold of the reaction, as it is implied by the values of binding energies in Table~\ref{tab:tableNCSMstate}:
The experimental Q-value is -0.192 MeV, whereas the energies of the ground states in our calculation give
a Q-value of -0.556 and -0.465 MeV for $N_{\rm max}$= 6 and 8, respectively. 

A feature which is not reproduced by our calculation is the second peak in the total cross section, which corresponds
to a resonance with positive parity and uncertain spin assignment ($\frac{7}{2})^+$, located at 17.493 MeV in the $^9$Be
spectrum. We can only speculate on the reasons of this deficiency in our calculation: The resonant peak in question is pronounced 
in the $^8$Be($\alpha$-$\alpha$)-$n$ decaying channel~\cite{Tilley2004}, which is not explicitly included in the present cluster expansion. At the same time it should be remembered that the present results lack the effect of  3N forces, which can have an impact on the peak structure of the cross section.
Another possible reason could be the insufficient inclusion of short-range correlations in the NCSM/RGM model space. This could be corrected by coupling NCSM eigenstates
of $^9$Be, that is by working within the NCSMC framework. While efforts are being devoted to fully extend the NCSMC to transfer reactions involving p-shell targets,  we are currently not yet in the position to apply this formalism to the study of the $^7$Li($d$,$p$)$^8$Li transfer reaction. 

One way to overcome the limitations of our present calculation is to follow a more phenomenological approach (\lq NCSM/RGM phenomenology\rq)  and correct the NCSM energies for the $d$, $^7$Li and $^8$Li clusters in such a way that the difference between $d+^7$Li and $p+^8$Li thresholds are reproduced with a desired level of accuracy. %
%\begin{figure}[t]
%\includegraphics[width=\columnwidth,keepaspectratio]{PLOTS/EPS_FIG/sigma_tot_dLi7gspLi8gs_Nmax8_RGM_lab_thresh_bis2.eps}%
%\caption{\label{sigma_tot_thresh} $^7$Li($d$,$p$)$^8$Li integrated cross section at the deuteron laboratory energies up to 2.25 MeV computed in NCSM/RGM approach. Model-space basis size is truncated at $N_{\rm max}$= 8. Experimental data are from Refs.~\cite{Parker1966,Mingay1979, Elwyn1982,Filippone1982}. Magenta solid line corresponds to a calculation in NCSM/RGM phenomenology approach (see text for details). }
%\end{figure}
%
Specifically,  we set the g.s.\ energy of the deuteron to its experimental value (2.2245 MeV), whereas do not adjust the g.s.\ of $^7$Li, but only correct the energy of its $\frac{1}{2}^-$
first excited state to match the measured excitation energy. We then modify the g.s.\ energy of $^8$Li in order to reproduce the experimental Q-value and shift the energies of the three excited states included in the model space to reproduce the corresponding experimental excitation energies. The adjusted $^7$Li and $^8$Li energies are specified in the last column of Table~\ref{tab:tableNCSMstate} and the resulting effect on the computed cross section is displayed in Fig.~\ref{sigma_tot}.
This simple readjustment brings the calculated total cross section in fairly good agreement with the measured one (see thick-dashed magenta line in Fig.~\ref{sigma_tot}). The position of the first peak is slightly overestimated and the trend of the cross section up to  2.3 MeV is qualitatively reproduced. The second peak at about 1 MeV above the $d+^7$Li threshold is missing also in the adjusted 
calculation, which is consistent with our hypothesis that this is dominated by a $^8$Be($\alpha$-$\alpha$)-$n$ decay mode.

\begin{figure}[t]
\includegraphics[width=\columnwidth,keepaspectratio]{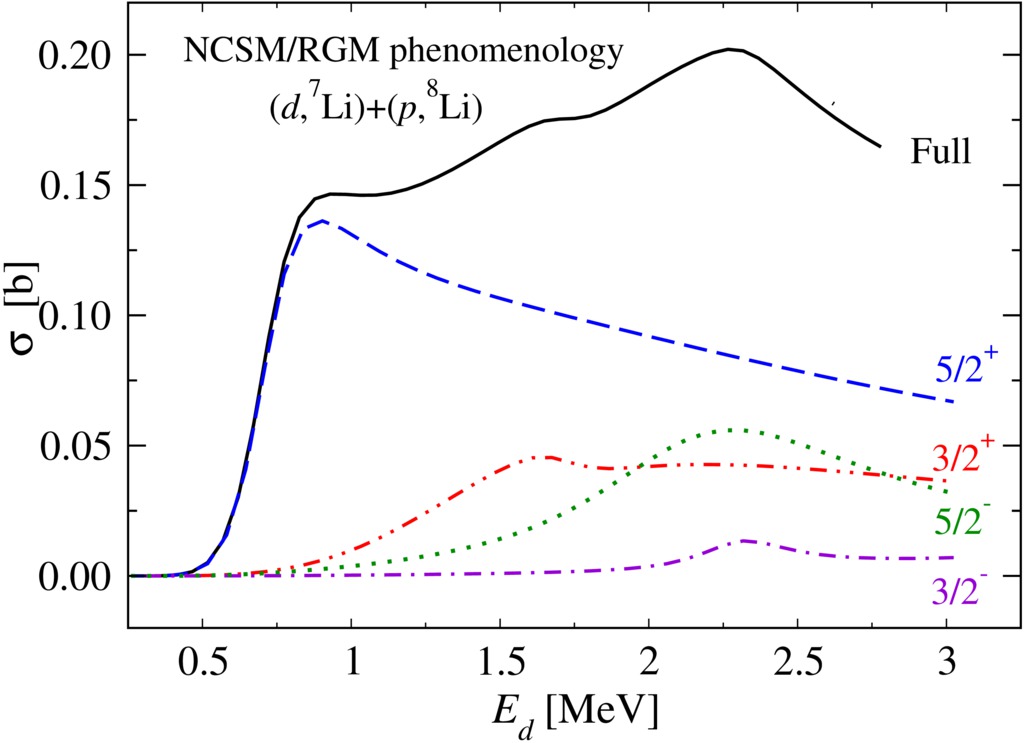}%
\caption{\label{sigma_tot_thresh_partials} Contribution of dominant $J^\pi$ $(T=\frac12)$ partial waves to the NCSM/RGM phenomenology total cross section (solid line) shown in Fig.~\ref{sigma_tot}.}
\end{figure}

Finally we study the contribution of the different partial waves to the total cross section of Fig.~\ref{sigma_tot} by repeating the calculation with only one component at the time. In this way we
are in the position to assess the impact of each partial wave on the cross section, and therefore to assign exactly spin, isospin and parity quantum numbers to the peaks appearing in the excitation function. The result of this analysis
is displayed in Fig.~\ref{sigma_tot_thresh_partials}, where one can see that the relevant contributions to the integrated cross section
is given by the partial waves with $J=\frac{3}{2}$ and $\frac{5}{2}$ and $T$=$\frac{1}{2}$. In particular, below 2 MeV in the deuteron kinetic energy the cross section is dominated by positive-parity partial waves while the impact of the 
negative-parity ones becomes significant at higher energies.
The dominant role played by the $\frac{5}{2}^+$ partial wave on the first peak of the total cross section
confirms the conclusions we have drawn by considering the phase and eigenphase shifts in Figs.~\ref{eigen_minus_plus}  and ~\ref{phase_shift_5half}, where the channels with $\frac{5}{2}^+\frac{1}{2}$ quantum numbers are the most relevant 
ones above the threshold of the reaction. For the range of deuteron energies above 2 MeV, we see that the constant rising of the cross section is dominated by the $\frac{5}{2}^-$ partial wave. 
The analysis of the partial wave contributions in Fig.~\ref{sigma_tot_thresh_partials} has been performed on the cross section obtained from the 
NCSM/RGM phenomenology approach, i.e.
for cluster eigenstate energies adjusted to reproduce the reaction threshold. However the results of this analysis concerning the impact of different partial waves are valid 
also for the calculation without any adjustment.

\section{\label{sec:conclusion}Conclusions}

The description of deuteron stripping to resonant states of the compound nucleus, i.e. the nucleus composed by the 
resonant cluster deuteron-target, is demanding in terms of the theoretical tools required both in the formalism and 
in the computation, also for phenomenological approaches which rely on model potentials~\cite{Mukhamedzhanov2011}.

In this paper we presented an extension of the \textit{ab initio} NCSM/RGM approach to $(d,p)$ transfer reactions with p-shell ($A>4$) targets. For this purpose we considered a microscopic Hamiltonian truncated at the NN level, and extended an efficient algorithm recently
introduced to study nucleon-nucleus collisions with the inclusion of 3N forces~\cite{Langhammer2015}. This algorithm significantly reduces the overhead in the computation of the transition density matrix elements and eliminates the necessity of storing them before the actual computation of the Hamiltonian kernels. 

We then applied the newly developed approach to the description of the ${}^{7}$Li($d$,$p$)${}^{8}$Li as well as to the $d-^7$Li and $p-^8$Li elastic scattering. 
For the case of the elastic process we also performed calculations within the NCSMC formalism, with an extended
basis including the NCSM state of $^9$Be, whereas the calculations for the transfer reaction were performed at the NCSM/RGM level
considering explicitly the ($p,^8$Li) mass partition in the nuclear wave function both at short distances and in the asymptotic part.

We discussed the experimental spin-parity assignments of the resonances of the compound $^9$Be nucleus,
especially for the resonance detected at deuteron energy of 0.78 MeV, the measured absolute yield of which is used
as a calibration for the target thickness in proton capture experiments on $^7$Be. We found that our calculations support a spin-parity assignment of $J^\pi=\frac{5}{2}^{+}$ for this resonance: This is at odds with the experimental assignment of $J^\pi$=$\frac{5}{2}^{-}$. In general, we showed the interplay between deuteron-$^7$Li and proton-$^8$Li channels in explaining some features of the $^9$Be spectrum.

Our conclusions need to be confirmed by further calculations: Indeed, owning to the non-completeness of the basis truncated at $N_{\rm max}=8$,
we cannot establish to which extent the disagreement with respect to the experimental data can be ascribed to the lack of convergence of the calculation,  rather than to missing 3N forces or
degrees of freedom in the model-space basis such as the $n$-$^8$Be channel. 

Our future developments in the line of the present work are directed to the development of the NCSMC for the \textit{ab initio} description of $(d,N)$ transfer reactions. More work is also necessary to enable the convergence of our calculation, in particular when heavier targets will be considered, and include the effect of 3N forces~\cite{Hupin2013,Hupin2015}.  The first goal can be achieved by working with cluster wave functions truncated according to the importance of the different components~\cite{Roth2007}, and the second by taking into account 3N forces in an effective way through a normal ordering approximation.

\appendix

\section{\label{App_gen}NCSM/RGM and NCSMC formalism}
In this Appendix we collect the main equations of the \textit{ab initio} NCSMC and NCSM/RGM formalism. Generally speaking, the 
former can be seen as an extension of the latter, particularly concerning the treatment of the correlations in the $A$-nucleons
wave function. The microscopic-cluster basis of the NCSM/RGM is complemented by NCSM states of the $A$-nucleons system in the NCSMC approach, producing a basis capable to describe long- and short-range correlations
on the same footing.

\subsection{\label{App_1}Coupled-channel NCSM/RGM equations}
In the Hilbert space spanned by the basis states $\hat{\mathcal A}_{\nu}\,|\Phi^{J^\pi T}_{\nu r}\rangle$ of Eq.~(\ref{eq:trial}), the many-body nuclear problem assumes the form of the set of coupled integral-differential equations,
\begin{equation}
\sum_{\nu}\int dr \,r^2\left[{\mathcal H}^{J^\pi T}_{\nu^\prime\nu}(r^\prime, r)-E\,{\mathcal N}^{J^\pi T}_{\nu^\prime\nu}(r^\prime,r)\right] \frac{g^{J^\pi T}_\nu(r)}{r} = 0\,,\label{RGMeq}
\end{equation}
where the norm kernel,
\begin{eqnarray}
{\mathcal N}^{J^\pi T}_{\nu^\prime\nu}(r^\prime, r) &=& \left\langle\Phi^{J^\pi T}_{\nu^\prime r^\prime}\right|\hat{\mathcal A}_{\nu^\prime}\hat{\mathcal A}_{\nu}\left|\Phi^{J^\pi T}_{\nu r}\right\rangle\,,\label{N-kernel}
\end{eqnarray}
results from the non-orthogonality of the basis states, and the Hamiltonian kernel
\begin{eqnarray}
{\mathcal H}^{J^\pi T}_{\nu^\prime\nu}(r^\prime, r) &=& \left\langle\Phi^{J^\pi T}_{\nu^\prime r^\prime}\right|\hat{\mathcal A}_{\nu^\prime}H\hat{\mathcal A}_{\nu}\left|\Phi^{J^\pi T}_{\nu r}\right\rangle\,,\label{H-kernel}
\end{eqnarray}
is given by the matrix elements of the internal $A$-nucleon microscopic Hamiltonian,
\begin{equation}\label{Hamiltonian}
H=T_{\rm rel}(r)+ {\mathcal V}_{\rm rel} +\bar{V}_{\rm C}(r)+H_{(A-a)}+H_{(a)}\,.
\end{equation}
The decomposition of the Hamiltonian of Eq.~(\ref{Hamiltonian}) contains the relative kinetic energy $T_{\rm rel}(r)$
and ${\mathcal V}_{\rm rel}$, which is the sum of all interactions between nucleons belonging to different clusters after subtraction of the average Coulomb interaction between them, explicitly singled out in the term $\bar{V}_{\rm C}(r)=Z_{1\nu}Z_{2\nu}e^2/r$, where $Z_{1\nu}$ and $Z_{2\nu}$ are the charge numbers of the clusters for a given channel $\nu$. In the present work we consider only the NN part of the nuclear interaction, then the intercluster interaction reads,
\begin{eqnarray}
{\mathcal V}_{\rm rel} &=& \sum_{i=1}^{A-a}\sum_{j=A-a+1}^AV_{ij}
-\bar{V}_{\rm C}(r),
\end{eqnarray}
with $V_{ij}$ being the NN strong and Coulomb part of the interaction between nucleons.

The key quantities that must be calculated by solving Eq.~(\ref{RGMeq}) in order to derive the scattering observables, such as phase shifts and cross section, are the relative motion amplitudes $g^{J^\pi T}_\nu(r)$ in the trial wave function~(\ref{eq:trial}). The solutions are obtained by means of the microscopic R-matrix method on a Lagrange mesh~\cite{Baye1986,Descouvemont2010}.

\subsection{\label{App_1bis}Coupled-channel NCSMC equations}

In the NCSMC approach, the NCSM/RGM ansatz for the $A$-body wave function of Eq.~(\ref{eq:trial}) is complemented with an expansion over square-integrable $A$-body basis states  according to: 
\begin{align}
\label{NCSMC_wav}
\!\!\!\ket{\Psi^{J^\pi T}_A} \!=\!  \sum_\lambda c_\lambda \ket{A \lambda J^\pi T} \! +\!  \sum_{\nu} \int dr \,r^2\frac{g^{J^\pi T}_{\nu}(r)}{r}\,\hat{\mathcal A}_{\nu}\,|\Phi^{J^\pi T}_{\nu r}\rangle\, ,
\end{align}
where the $\ket{A \lambda J^\pi T} $ states are NCSM energy eigenstates expanded over a set of antisymmetric $A$-nucleon HO basis states containing up to  %truncated by 
$N_{\rm max}$ %, the maximum %al  
%allowed 
 HO excitations above the lowest possible configuration. They are obtained by diagonalizing the intrinsic Hamiltonian, $\hat{H}=\hat{T}_{\rm int}+\hat{V}$, %of the $A$-nucleon system, 
\begin{equation}\label{NCSM_eq}
\hat{H} \ket{A \lambda J^\pi T} = E_\lambda \ket{A \lambda J^\pi T} \; ,
\end{equation}
%
%with $\hat{H}=\hat{T}_{\rm int}+\hat{V}$, %we obtain the energy eigenstates expanded in the $N_{\rm max}\hbar\Omega$ basis, $\ket{ANiJ^\pi T}$. %The 
%Here, 
where $\hat{T}_{\rm int}$ is the internal kinetic energy operator and %the 
$\hat{V}$ the NN or NN+3N interaction.

With the ansatz of Eq.~(\ref{NCSMC_wav}) we must solve the $A$-nucleons problem for two types of unknowns, the discrete, $c_\lambda$, and the continuous, $g^{J^\pi T}_{\nu}(r)$. The coupled-channel generalized Schr\"odinger equation~(\ref{RGMeq}) becomes

\begin{eqnarray}\label{eq:formalism_110}
  \left(
     \begin{array}{cc}
        H_{NCSM} & \bar{h} \\
        \bar{h}  & \overline{\mathcal{H}} 
     \end{array}
  \right)
  \left(
     \begin{array}{c}
          c \\
		  {\chi}
     \end{array}
  \right)
  = 
  E
  \left(
     \begin{array}{cc}
        1 & \bar{g} \\
        \bar{g}  & 1 
     \end{array}
  \right)
  \left(
     \begin{array}{c}
          c \\ 
		  {\chi}
     \end{array}
  \right),
\end{eqnarray}
where $\chi_{\nu} (r)$ 
are the relative wave functions in the NCSM/RGM sector when working with the orthogonalized cluster channel states~\cite{Quaglioni2009}. $H_{NCSM}$ denotes the diagonal matrix of the NCSM energy eigenvalues, which give the NCSM sector of the Hamiltonian kernel, while 
$\overline{\mathcal{H}}$ is the orthogonalized NCSM/RGM kernel~\cite{Quaglioni2009}.
The coupling between the two sectors is described by the 
overlap, $\bar{g}_{\lambda \nu}(r)$ (not to be confused with the relative motion function $g^{J^\pi T}_\nu(r)$), and Hamiltonian, $\bar{h}_{\lambda \nu}(r)$, form factors respectively  proportional to the $\braket{A \lambda J^\pi T}{\hat{\mathcal{A}}_{\nu} \Phi_{\nu r}^{J^\pi T  }}$ and 
$\matrEL{A \lambda J^\pi T} {\hat{H} \hat{\mathcal{A}}_{\nu}} {\Phi_{\nu r}^{J^\pi T  }}$ matrix elements.

\section{\label{App_2}Hamiltonian kernels for deuteron-induced reactions}
For the sake of completeness, in this Appendix we summarize the Hamiltonian kernels which are used for the description of deuteron-induced reactions in the NCSM/RGM framework, for both elastic and single-nucleon transfer reactions. 

\subsection{\label{app_2:subsec1}Hamiltonian kernels for ($A$-2,2) mass partition}

For identical ($A$-2,2) mass partitions in both initial and final channels, the Hamiltonian kernel can be cast in the form,
\begin{widetext}
\begin{eqnarray}
\mathcal{H}_{\nu'\nu}^{J^\pi T}(r',r)&=&
\left<\Phi_{\nu' r'}^{J^\pi T}\right|\hat{\mathcal{A}}_{(A-2,2)}H\hat{\mathcal{A}}_{(A-2,2)}
\left|\Phi_{\nu r}^{J^\pi T}\right>\notag
 =\left<\Phi_{\nu' r'}^{J^\pi T}\right|H\hat{\mathcal{A}}^2_{(A-2,2)}
\left|\Phi_{\nu r}^{J^\pi T}\right>\notag\\
&=&\left[{T}_{\rm rel}(r')+\bar{V}_C(r')+E_{\alpha_1'}^{I_1'T_1'} +E_{\alpha_2'}^{I_2'T_2'}\right]
\mathcal{N}_{\nu'\nu}^{J^\pi T}(r', r)+\mathcal{V}^{J^\pi T}_{\nu' \nu}(r',r),
\end{eqnarray}
where ${T}_{\rm rel}(r')$ and $\bar{V}_C(r')$ are defined in Eq.~(\ref{Hamiltonian}), $E_{\alpha_1'}^{I_1'T_1'}$ and $E_{\alpha_2'}^{I_2'T_2'}$ are NCSM energy eigenvalues for the two clusters, and the potential kernel is given by
\begin{subequations}
\label{V-kernel}\begin{align}
\mathcal{V}^{J^\pi T}_{\nu' \nu}(r',r) &= 
\left<\Phi_{\nu' r'}^{J^\pi T}\right|\mathcal{V}_{\rm rel}\hat{\mathcal{A}}^2_{(A-2,2)} \left|\Phi_{\nu r}^{J^\pi T}\right> 
  \\
& =  \sum_{n^\prime n} R_{n^\prime \ell^\prime} (r^\prime) R_{n \ell} (r)\,  \left[ 2(A-2) \left<\Phi_{\nu' n'}^{J^\pi T}\right| V_{A-2,A-1}(1-\hat P_{A-2,A-1})\left|\Phi_{\nu n}^{J^\pi T}\right>  \right. \\
& \quad\phantom{ =\sum_{n^\prime n} R_{n^\prime \ell^\prime} (r^\prime) R_{n \ell} (r)}
- 2(A-2) \left<\Phi_{\nu' n'}^{J^\pi T}\right| V_{A-2,A}\hat P_{A-2,A-1}\left|\Phi_{\nu n}^{J^\pi T}\right>  \\
& \quad\phantom{ =\sum_{n^\prime n} R_{n^\prime \ell^\prime} (r^\prime) R_{n \ell} (r)}  
- 2(A-2)(A-3) \left<\Phi_{\nu' n'}^{J^\pi T}\right| V_{A-3,A}(1-\hat P_{A-3,A}) \hat P_{A-2,A-1}\left|\Phi_{\nu n}^{J^\pi T}\right> \\
&\quad\phantom{ =\sum_{n^\prime n} R_{n^\prime \ell^\prime} (r^\prime) R_{n \ell} (r)}
- 2(A-2)(A-3) \left<\Phi_{\nu' n'}^{J^\pi T}\right| V_{A-3,A-1}\hat P_{A-2,A-1}\left|\Phi_{\nu n}^{J^\pi T}\right> \\
&\quad\phantom{ =\sum_{n^\prime n} R_{n^\prime \ell^\prime} (r^\prime) R_{n \ell} (r)}
\left. + (A-2)(A-3)(A-4) \left<\Phi_{\nu' n'}^{J^\pi T}\right| V_{A,A-4}(1-\hat P_{A-2,A-1}) \hat P_{A-3,A}\left|\Phi_{\nu n}^{J^\pi T}\right> \right]\,.
\label{antisym_HAm2}
\end{align}
\end{subequations}
\end{widetext}

Among the five separate terms in Eq.~(\ref{V-kernel}) we are interested in particular in the one of line~(\ref{antisym_HAm2}), depending on the three-body density of the target nucleus, that is a challenge in terms of the computational resources needed to compute it, due to their rapidly increasing number in the multi-major shell basis spaces. Its representation with respect 
to the SD channel states of Eq.~(\ref{SD-basis-ab}) is (see Eq. (24) in Ref.\cite{Navratil2011}),

\begin{widetext}
\begin{align}
 &\leftsub{\rm SD}{\left\langle\Phi^{J^\pi T}_{\kappa_{ab}^\prime}\left|V_{A,A-4} \hat{P}_{A-2,A-1}\hat{P}_{A-3,A} \right|\Phi^{J^\pi T}_{\kappa_{ab}}\right\rangle}_{\rm  SD} 
 \nonumber \\[2mm]
 & \qquad =
\frac{1}{2}\frac{1}{(A-2)(A-3)(A-4)}
\sum
\left\{ \begin{array}{@{\!~}c@{\!~}c@{\!~}c@{\!~}} 
J_{de} & j_b^\prime & X \\[2mm] 
I^\prime & j_e^\prime & j_a^\prime 
\end{array}\right\}
\left\{ \begin{array}{@{\!~}c@{\!~}c@{\!~}c@{\!~}} 
T_{de} & \frac{1}{2} & \tau_X \\[2mm] 
T_2^\prime & \frac{1}{2} & \frac{1}{2} 
\end{array}\right\}
\left\{ \begin{array}{@{\!~}c@{\!~}c@{\!~}c@{\!~}} 
I_1 & I & J \\[2mm] 
X & j_e^\prime & I^\prime \\[2mm] 
I_\beta & Y & I_1^\prime 
\end{array}\right\}
\left\{ \begin{array}{@{\!~}c@{\!~}c@{\!~}c@{\!~}} 
T_1 & T_2 & T \\[2mm] 
\tau_X & \frac{1}{2} & T_2^\prime \\[2mm] 
T_\beta & \tau_Y & T_1^\prime 
\end{array}\right\}
\nonumber \\[2mm]
& \qquad \phantom{=} \times
\hat{I}^\prime \, \hat{X} \, \hat{Y} \, \hat{J}_{de} \, \hat{T}_2^\prime \, \hat{\tau}_X \, \hat{\tau}_Y \, \hat{T}_{de} \,
(-1)^{I^\prime+J_{de}+J-I_1^\prime+j_e-j_d} \; (-1)^{T_2^\prime+T_{de}+T-T_1^\prime}
\nonumber \\[2mm]
&\qquad \phantom{=} \times
 \leftsub{\rm SD}{\left\langle A{-}2\, \alpha_1^\prime I_1^\prime T_1^\prime \left | \left |\left | 
\left((a^\dagger_a a^\dagger_b)^{(I T_2)} a^\dagger_{e^\prime}\right)^{(Y \tau_Y)}\right | \right | \right | A{-}5 \,\beta I_\beta T_\beta\right\rangle}_{\rm SD}
\nonumber \\[2mm]
&\qquad \phantom{=} \times
 \leftsub{\rm SD}{\left\langle  A{-}5 \,\beta I_\beta T_\beta \left |\left |\left | \left(\tilde{a}_{b^\prime} (\tilde{a}_{e}\tilde{a}_{d})^{(J_{de} T_{de})}\right)^{(X\tau_x)}
\right | \right | \right | A{-}2\, \alpha_1 I_1 T_1\right\rangle}_{\rm SD}
\nonumber \\[2mm]
&\qquad \phantom{=} \times
\sqrt{1+\delta_{a^\prime,e^\prime}} \, \sqrt{1+\delta_{d,e}}\;\;
\left\langle a^\prime e^\prime J_{de} T_{de} \left| V \right| d \,e\, J_{de} T_{de} \right\rangle 
\,,
\label{V-kernel_5-compl}
\end{align}
\end{widetext}
where the sum runs over the quantum numbers $\beta, I_\beta, T_\beta$, $d \equiv n_d \ell_d j_d \tfrac12$ etc., $e, e^\prime$, $J_{de}, T_{de}$, $X, Y, \tau_X$, and $\tau_Y$.  Eq.~(\ref{V-kernel_5-compl}) can be compared to the much more compact and more practical expression~(\ref{Summary_II}).

\subsection{\label{app_2:subsec2}Hamiltonian coupling kernels for ($A$-2,2)-($A$-1,1) mass partitions}

The contributions to the Hamiltonian  kernels coming from the off-diagonal matrix elements between the two mass partitions $(A-1,1)$ and $(A-2,2)$ appear in the many-body generalized Schr\"odinger equation when both deuterium-nucleus and nucleon-nucleus are included in the cluster expansion channels. For the NN part of the nuclear interaction they are given by

\begin{widetext}
\begin{subequations}
	 \label{pot-coupled}
 \begin{align}
	\bar {\mathcal V}^{J^\pi T}_{\nu^\prime\nu}(r^\prime, r) = &
	 \sqrt{\tfrac{A-1}2} \sum_{n^\prime\,n} R_{n^\prime\ell^\prime}(r^\prime) R_{n\ell}(r) \left[ 2(A-2) \left\langle\Phi^{J^\pi T}_{\nu^\prime n^\prime}\right| V_{A-2,A}(1-\hat P_{A-2,A})\left|\Phi^{J^\pi T}_{\nu n}\right\rangle \right.  \\[2mm]
	&\quad  + \left\langle\Phi^{J^\pi T}_{\nu^\prime n^\prime}\right| V_{A-1,A}\left|\Phi^{J^\pi T}_{\nu n}\right\rangle 
	+ (A-2) \left\langle\Phi^{J^\pi T}_{\nu^\prime n^\prime}\right| V_{A-2,A-1}\left|\Phi^{J^\pi T}_{\nu n}\right\rangle \\[2mm]
	 &\quad \left. -(A-2)(A-3) \left\langle\Phi^{J^\pi T}_{\nu^\prime n^\prime}\right| \tfrac 12 \hat P_{A-2,A}V_{A-3,A-2} + V_{A-3,A-2} \hat P_{A-2,A} \left|\Phi^{J^\pi T}_{\nu n}\right\rangle \right ] \,.\label{pot-coupled-last}
\end{align} 
\end{subequations}
\end{widetext}
The last two terms in Eq.~(\ref{pot-coupled}) are written in terms of matrix elements of two creation and three annihilation field operators acting on the target wave functions, as explicitly shown in Eq.~(\ref{Summary_I_bis_total}).  

%\bibliography{d_p_transfer_10}% Produces the bibliography via BibTeX.

\newpage

\end{document}